\documentclass[a4paper,11pt]{article}
\usepackage{jheppub}
\usepackage{epsfig}
\usepackage{amsmath}
\usepackage{graphicx}
\usepackage {float}
\usepackage[T1]{fontenc}
\usepackage[utf8]{inputenc}

\def\A0#1{\Pi_{\rm #1}(0)}
\def\AP0#1{\Pi'_{\rm #1}(0)}
\def\be{\begin{equation}}
\def\ee{\end{equation}}
\def\bea{\begin{array}}
\def\eea{\end{array}}
\def\beqa{\begin{eqnarray}}
\def\eeqa{\end{eqnarray}}
\def\beqas{\begin{eqnarray*}}
\def\eeqas{\end{eqnarray*}}

\def\bp{\begin{picture}}
\def\ep{\end{picture}}
\def\bc{\begin{center}}
\def\ec{\end{center}}
\def\bfig{\begin{figure}}
\def\efig{\end{figure}}

\def\bit{\begin{itemize}}
\def\eit{\end{itemize}}
\def\nn{\nonumber}
\def\f{\frac}

\def\[{\left[}
\def\]{\right]}
\def\({\left(}
\def\){\right)}

\def\..{\left.}
\def\.{\right.}
\def\tl{\tilde}
\def\ra{\rightarrow}
\def\la{\leftarrow}

\def\tm{\times}

\def\da{\dagger}

\def\la{\lambda}

\def\al{\alpha}

\def\ka{\kappa}

\def\ep{\epsilon}

\def\pa{\partial}
\def\pr{\prime}

\title{Interplay between the muon $g-2$ anomaly and the PTA nHZ gravitational waves from domain walls in next-to minimal supersymmetric standard model}
\author[a]{Ming Xia Huang,}
\author[a]{Fei Wang\footnote{Corresponding author},}
\author[a]{Ying Kai Zhang}
\affiliation[a]{School of Physics and Laboratory of Zhongyuan Light, Zhengzhou University, Zhengzhou 450000, P. R. China}
\emailAdd{feiwang@zzu.edu.cn}
\abstract{With some explicitly $Z_3$ breaking terms in the NMSSM effective superpotential and scalar potential, domain walls (DWs) from spontaneously breaking of the discrete symmetry in approximate $Z_3$-invariant NMSSM can collapse and lead to observable stochastic gravitational wave (GW) background signals. In the presence of a hidden sector, such terms may originate from the geometric superconformal breaking with holomorphic quadratic correction to frame function when the global scale-invariant superpotential is naturally embedded into the canonical superconformal supergravity models. The smallness of such mass parameters in the NMSSM may be traced back to the original superconformal invariance. Naive estimations indicate that a SUSY explanation to muon $g-2$ anomaly can have tension with the
constraints on SUSY by PTA data, because large SUSY contributions to $\Delta a_\mu$ in general
needs relatively light superpartners while present $\Omega_{gw}^0$
can set the lower bounds for $m_{soft}$.  We calculate numerically the signatures of GW produced from the collapse of DWs and find that the observed nHZ stochastic GW background by NANOGrav, etc., can indeed be explained with proper tiny values of $\chi m_{3/2}\sim 10^{-14}{\rm  eV}$ for $\chi S^2$ case (and $\chi m_{3/2}\sim 10^{-10}{\rm  eV}$ for $\chi H_u H_d$ case), respectively. Besides, there are still some parameter points, whose GW spectra intersect with the NANOGrav signal region, that can explain the muon $g-2$ anomaly to $1\sigma$ range.}

\begin{document}
\maketitle \indent
\newpage
\section{\label{sec-1}Introduction}
Gravitational waves(GWs) can be produced in the early Universe, forming a stochastic GW background that can possibly be detected via radio telescopes using the pulsar timing arrays (PTAs) when the frequency lies around $10^{-8}$ Hz.
 Recently, various PTA collaborations, including the NANOGrav~\cite{NANOGrav:2023gor,NANOGrav:2023hde,NANOGrav:2023hvm}, EPTA~\cite{Antoniadis:2023ott},
CPTA~\cite{Xu:2023wog} and PPTA~\cite{Reardon:2023gzh,Reardon:2023zen,Zic:2023gta}, have reported evidences for a stochastic GW background
in the nHz frequency band. Many models have been proposed to explain this signal and potential sources for such a stochastic
 GW background involve supermassive black hole binaries~\cite{Bi:2023tib, Bian:2023dnv, Ellis:2023dgf, Deng:2023btv, Shen:2023pan, Zhang:2023lzt, Cannizzaro:2023mgc, Barausse:2023yrx, Aghaie:2023lan, Ellis:2023oxs},~primordial black hole~\cite{Gouttenoire:2023nzr,Unal:2023srk, Franciolini:2023pbf, Inomata:2023zup, Depta:2023qst, Bhaumik:2023wmw}, primary perturbation~\cite{Anchordoqui:2023tln, Liu:2023ymk, Basilakos:2023xof, Jin:2023wri, Gorji:2023sil, Harigaya:2023pmw}, inflation~\cite{Borah:2023sbc,Murai:2023gkv, Datta:2023vbs, Niu:2023bsr, Vagnozzi:2023lwo, Chowdhury:2023opo, Firouzjahi:2023lzg, Choudhury:2023kam, HosseiniMansoori:2023mqh, Cheung:2023ihl, Bousder:2023ida, Jiang:2023gfe, Zhu:2023lbf, Yi:2023npi, Fu:2023aab},
  scalar induced gravitational waves~\cite{Cai:2023dls, Wang:2023ost, Yi:2023mbm, Zhu:2023faa, You:2023rmn, Balaji:2023ehk, Liu:2023pau, Yi:2023tdk, Frosina:2023nxu, Choudhury:2023wrm}, cosmic string~\cite{Wang:2023len, Ellis:2023tsl, Kitajima:2023vre, Servant:2023mwt, Antusch:2023zjk,  Ahmed:2023rky, Ahmed:2023pjl}, domian walls~\cite{Guo:2023hyp, Lu:2023mcz, Kitajima:2023cek, Lazarides:2023ksx, Blasi:2023sej, Gouttenoire:2023ftk, Barman:2023fad, Babichev:2023pbf, Gelmini:2023kvo, Ge:2023rce, Zhang:2023nrs}, first-order phase transitions~\cite{Zu:2023olm, Li:2023yaj, Athron:2023mer, Xiao:2023dbb, Li:2023bxy, Ghosh:2023aum, Han:2023olf, Fujikura:2023lkn, Addazi:2023jvg, Jiang:2023qbm, Abe:2023yrw, DiBari:2023upq, Gouttenoire:2023bqy, Salvio:2023ynn, Ahmadvand:2023lpp, Wang:2023bbc, An:2023jxf}  and other astrophysical/cosmological  GWs or constraints \cite{Figueroa:2023zhu, Wu:2023hsa, Lambiase:2023pxd, Yang:2023aak,  Franciolini:2023wjm, Oikonomou:2023qfz, Du:2023qvj, Ye:2023xyr, Ben-Dayan:2023lwd, Maji:2023fhv, InternationalPulsarTimingArray:2023mzf, Ding:2023xeg, Cyr:2023pgw, Geller:2023shn}.
Among all beyond the Standard Model (SM) fits to NANOGrav results, the domain walls (DWs) and phase transition sources perform better.  So, it is well motivated to concentrate on the possibility that DWs from new physics beyond the SM act as possible cosmological sources of nHZ stochastic GWs. DWs are sheetlike topological defects generated from the spontaneous breaking of discrete symmetry, which can carry useful information related to the underlying theory. We would like to see if the recent nHz GW background data can shed new lights on the relevant new physics models that generate the DWs, for example, the popular $Z_3$-invariant Next-to-Minimal Supersymmetric Standard Model~(NMSSM)~\cite{Ellwanger:2009dp,Maniatis:2009re} of low-energy supersymmetry (SUSY).

  Low-energy SUSY is the most promising new physics beyond the SM of particle physics. The SUSY framework can solve almost all the theoretical and aesthetic problems that bother the SM. In particular, the~discovered 125 GeV Higgs scalar lies miraculously in the {small} {`}$115-135${'} GeV window predicted by the low-energy SUSY, which can be seen as a strong hint of the existence of low-energy SUSY. NMSSM can elegantly solve the $\mu$ problem of Minimal Supersymmetric Standard Model~(MSSM) with an additional singlet sector. In addition, with~additional tree-level contributions or through doublet-singlet mixing, NMSSM can accommodate easily the discovered 125 GeV Higgs boson mass with low electroweak fine tuning. The $Z_3$-invariant NMSSM is the most predictive version of NMSSM realizations with less new input parameters. However, the VEVs of Higgs fields $S,H_u,H_d$ at the electroweak phase transition era will also trigger the breaking of the $Z_3$ discrete symmetry, potentially generated cosmologically problematic DWs because the DWs energy density decreases slower than radiation and matter, and would soon dominate the total energy density of the universe. As long as the discrete symmetry is exact, such topological configurations are stable. However, it was argued that DWs with a tension as large as $\sigma > {\cal O}({\rm MeV}^3)$ must not exist in the present universe, which is known as Zel'dovich-Kobzarev-Okun bound~\cite{ZKO-bound}.
  Fortunately, the degeneracy of the minima from the discrete symmetry can be lifted by an energy bias in the potential, causing the collapse of DWs at the early time of the Universe and produce large amounts of GWs. Such GWs can form a stochastic background that persist to the present Universe, potentially being probed by recent PTA experiments.

  There are several terms in the general NMSSM superpotential that can explicitly violate the $Z_3$ discrete symmetry to trigger the decay of DWs~\cite{1503.06998,Allanach:2008qq}. We argue that the smallness of the quadratic terms may be traced back to the original superconformal invariance for the matter part of the supergravity action when the global scale-free NMSSM is embedded to supergravity. Unlike the pure supergravity part, which breaks superconformal symmetry after the gauge fixing, the matter part alone remains superconformal. Additional real parts of the holomorphic functions in the frame function are adopted to break the superconformal symmetry, which result in the scale-invariance violation quadratic terms (such as $H_u H_d$ or $S^2$ terms) in the superpotential in the presence of a hidden sector.

    It is well known that the theoretical prediction of the muon anomalous magnetic moment $a_\mu\equiv (g-2)_\mu/2$ for SM has subtle deviations from the experimental values. In fact, combining the recent reported FNAL muon $g-2$ measurement with the previous BNL+FNAL results~\cite{g-2:BNL,g-2:PDG,g-2:FNAL}, the~updated world average experimental value of $a_\mu$~\cite{g-2:FNAL2} has a $5.1\sigma$ deviation from the SM predictions~\cite{g-2:Th} with
 \beqa
\Delta a^{\rm{FNAL+BNL}}_\mu =(24.9 \pm  4.8)  \tm 10^{-10}~,
\eeqa
imposing very stringent constraints on various new physics models, including the $Z_3$-invariant NMSSM. So, it is fairly interesting to see if the approximate $Z_3$-invariant NMSSM (with tiny explicit discrete symmetry breaking terms) can consistently lead to the signals of GWs by the collapse of DWs in the parameter regions allowed by the muon $g-2$ explanation and collider data. Such a survey is fairly nontrivial, as large SUSY contributions to $\Delta a_\mu$ always prefer $m_{soft}\sim {\cal O}(10^2){\rm GeV}$ while recent NANOGrav data can lead to approximate low bounds $m_{soft}\gtrsim {\cal O}(1){\rm TeV}$.

This paper is organized as follows. In Sec~\ref{sec-2}, we discuss the formation of DWs from the spontaneous breaking of $Z_3$ discrete symmetry. In Sec~\ref{sec-3}, the collapse of DWs from explicitly  $Z_3$-violation terms are discussed. Such small $Z_3$ violation terms can be the consequence of superconformal violation in the matter part of supergravity in the presence of a hidden sector.  In Sec~\ref{sec-4}, the GWs signals related to the collapse of DWs are discussed. The interplay between the muon $g-2$ anomaly and the PTA nHZ GWs signals from DWs are also discussed. Sec~\ref{sec:conclusions} contains our conclusions.

\section{\label{sec-2}DWs from $Z_3$-invariant NMSSM}
As noted previously, NMSSM is well motivated theoretically to solve the $\mu$ problem. A bare $\mu$ term $\mu H_u H_d$ in the NMSSM superpotential is forbidden if we impose a discrete $Z_3$ symmetry, under which the Higgs superfields $S, H_u, H_d$ transform nontrivially. The $Z_3$ invariant superpotential couplings are given by~\cite{Ellwanger:2009dp,Maniatis:2009re}
\beqa
W_{Z_3 NMSSM}&=&W_{MSSM}|_{\mu=0}+\la S H_u H_d+\f{\ka}{3} S^3~,~
\eeqa
with
\beqa
W_{MSSM}|_{\mu=0}=y^u_{ij} Q_{L,i} H_u U_{L,j}^c- y^d_{ij} Q_{L,i} H_d D_{L,j}^c-y^e_{ij} L_{L,i} H_d E_{L,j}^c~.
\eeqa
 {The} soft SUSY breaking parameters can be given as
\beqa
{\cal L}_{\tiny Z_3NMSSM}^{soft}={\cal L}_{MSSM}^{soft}|_{B=0}-\(A_\la \la S H_u H_d+A_\ka \f{\ka}{3} S^3\)-m_S^2 |S|^2~,
\eeqa
by proper SUSY breaking mechanisms, such as AMSB~\cite{Fei:1602.01699} and Yukawa mediation mechanisms~\cite{Wang:2015apa, Fei:1703.10894}. From the superpotential and the soft SUSY breaking terms, one obtains the Higgs potential
\beqa
V & = & \left|\lambda \left(H_u^+ H_d^- - H_u^0
H_d^0\right) + \kappa S^2 \right|^2 \nn \\
&&+\left(m_{H_u}^2 + \left|\lambda S\right|^2\right)
\left(\left|H_u^0\right|^2 + \left|H_u^+\right|^2\right)
+\left(m_{H_d}^2 + \left|\lambda S\right|^2\right)
\left(\left|H_d^0\right|^2 + \left|H_d^-\right|^2\right) \nn \\
&&+\frac{g_1^2+g_2^2}{8}\left(\left|H_u^0\right|^2 +
\left|H_u^+\right|^2 - \left|H_d^0\right|^2 -
\left|H_d^-\right|^2\right)^2
+\frac{g_2^2}{2}\left|H_u^+ H_d^{0*} + H_u^0 H_d^{-*}\right|^2\nn \\
&&+m_{S}^2 |S|^2
+\big( \lambda A_\lambda \left(H_u^+ H_d^- - H_u^0 H_d^0\right) S +
\frac{1}{3} \kappa A_\kappa\, S^3 + \mathrm{h.c.}\big),
\label{2.9e}
\eeqa
where $g_1$ and $g_2$ denote the $U(1)_Y$ and $SU(2)_L$ gauge couplings,
respectively. Electroweak symmetry breaking~(EWSB) are triggered by the VEVs $v_u\equiv\langle H_u\rangle$, $v_d\equiv\langle H_d\rangle$ while the effective $\mu_{eff}$ parameter
\beqa
\mu_{eff} \equiv \lambda s,
\eeqa
is generated by the VEV $s\equiv\left< S\right>$. Such Higgs VEVs also break spontaneously the $Z_3$ discrete symmetry and lead to the formation of DWs.

As the $Z_3$-NMSSM scalar potential is invariant under the $Z_3$ transformation, the scalar potential  has three degenerate minima that can be parametrized by $(|s|\omega^i,|v_u|\omega^i,|v_d|\omega^i)$ for $(i=0,1,2)$ with $\omega^3=1$. We use the conclusion in~\cite{NMSSM:SCPV} that the true minimum of the potential does not spontaneously break CP as the VEVs can always be made real by an appropriate field redefinition, up to the existence of
three degenerate vacua related to each other by $Z_3$ transformations.

  DWs are located around boundaries of these three vacua.
A planar domain wall solution perpendicular to the $z$-axis can be found by solving numerically the equation of motions for $S,H_u,H_d$ fields with the boundary conditions
\beqa
\lim\limits_{z\ra-\infty}\(S(z),H_u(z),H_d(z)\)&\longrightarrow& (|s|\omega^i,|v_u|\omega^i,|v_d|\omega^i)~,\nn\\
\lim\limits_{z\ra \infty}\(S(z),H_u(z),H_d(z)\)&\longrightarrow& (|s|\omega^{i+1},|v_u|\omega^{i+1},|v_d|\omega^{i+1})~,
\eeqa
such that the configuration interpolates smoothly between two vacua at $z\ra \pm\infty$.

 We adopt linearized interpolation curves for the two asymptotic domains as our initial guess of the solution and iterate these procedures until the solution converges. From the configuration of the DWs, the spatial distribution of the energy density can be calculated by
\beqa
\rho_{w}(z)=\sum\limits_{\phi_i}|\nabla \phi_i|^2+V_F+V_D+V_{soft}-V(s,v_u,v_d)~,\nn\\
=\sum\limits_{\phi_i=S,H_u,H_d}\left|\f{\pa \phi_i(z)}{\pa z}\right|^2+V(\phi_i)-V(s,v_u,v_d)~,
\eeqa
in which the boundary energy constant is subtracted such that $\rho_w\ra 0$ is satisfied for $z\ra \pm \infty$. The tension of the DWs, which
measures the energy stored per unit area on the wall, can be calculated from the energy-momentum tensor of the static solution. The calculation can be reduces to the integration of the energy density along the $z$-axis
\beqa
\sigma=\int\limits_{-\infty}^{\infty} dz \rho_w(z)~.
\eeqa
We show in Table\ref{tab1} the tension and energy density of DWs from $Z_3$-NMSSM for some benchmark points.
\begin{table}[H]
\centering
\begin{tabular}{|c|c|c|c|c|c|c|}
\hline
$tan\beta$ & $\lambda$ & $\kappa$ & $\text{A}_{\lambda}$ & $\text{A}_{\kappa}$ & $\mu_{\text{eff}}$  \\
\hline
45.9079 & 0.0166 & 0.0031 & 697.4122 & -291.0022 & 882.0991  \\
\hline
\end{tabular}
\begin{tabular}{|c|c|c|c|}
\hline
Particle & mass[GeV]& Particle &mass[GeV] \\\hline
$\tilde g$ &10585.1482&$\tilde e_L $&245.428730\\
$h_{1}$ &125.451953&$\tilde e_R $&346.932050\\
$h_{2}$ &252.599429&$\tilde \nu_{e_L} $&233.298454\\
$h_{3}$ &5918.80624&$\tilde \mu_L $&245.428730\\
$H^+$ &5919.72521&$\tilde \mu_R $&346.932050\\
$a$   &382.247468&$\tilde t_1 $ &10072.4644\\
$A$   &5918.80456&$\tilde t_2 $ &10188.7595\\
$\tilde d_L $&10303.7859&$\tilde \nu_{\mu_L} $&233.298454\\
$\tilde d_R $&10303.6597&$\tilde \tau_1 $&2573.50938\\
$\tilde u_L $&10303.5041&$\tilde \tau_2 $&2978.15121\\
$\tilde u_R $&10303.5716&$\tilde \nu_{\tau_L}$&2572.83541\\
$\tilde s_L $&10303.7859&$\tilde \chi_1 $&116.573893\\
$\tilde s_R $&10303.6597&$\tilde \chi_2 $&-290.383594\\
$\tilde c_L $&10303.5041&$\tilde \chi_3 $&344.081650\\
$\tilde c_R $&10303.5716&$\tilde \chi_4 $&-911.211934\\
$\tilde b_1 $&10184.2229&$\tilde \chi_5 $&911.837634\\
$\tilde b_2 $&10189.2163&$\tilde \chi^{\pm}_1$&116.576768\\
$M_W $&80.4200000&$\tilde \chi^{\pm}_2$&914.030425\\
$M_B $&4.18000000&$M_t$&173.400000\\
$M_Z $&91.1870000&$M_\tau$&1.77699995\\
\hline\hline
$\sigma$ & $7.3394\times 10^{11} \, \mathrm{GeV}^{3}$ & $t_{dec}$ & $3.6981 \times 10^{21} \, \mathrm{GeV}^{-1}$ \\
\hline
$\Omega_{GW} h^2$ & $2.1249 \times 10^{-9}$ & $f$ & $1.9025 \, \rm nHz$\\

\hline\hline
\end{tabular}
\label{tab1}
\caption{The tension, energy density of DWs from $Z_3$-invariant NMSSM and the SUSY contributions to $\Delta a_\mu$ for some benchmark points that can also satisfy all the low-energy collider constraints.}
\end{table}

We know that the validity of perturbation theory up to the GUT scale implies $\la<0.7\sim 0.8$~\cite{Ellwanger:2009dp} at the weak scale. Besides, the EWSB condition requires that
\beqa
\f{m_Z^2}{2}=|\la s|^2+m_{H_u}^2+\f{2}{\tan^2\beta}\(m_{H_d}^2-m_{H_u}^2 \)+{\cal O}(1/{\tan^4\beta})~.
\label{EWSB}
\eeqa
So, the singlet VEV $s$ always takes values of order the soft SUSY breaking scale, which is always much larger than the EW scale VEV $v_u$ and $v_d$ and act as the dominant spontaneously $Z_3$ discrete symmetry breaking source. In this case, as noted in~\cite{1503.06998}, the dominant contribution to the energy density and the tension of DWs comes from the singlet relevant terms in the scalar potential. From the equation of motion for the corresponding field, we can estimate the gradient of the field as
\beqa
\(\[\phi^\pr(z)\]^2\)^\pr=\f{\pa V(\phi)}{\pa \phi}\phi^\pr \Rightarrow \phi^\pr(z) \sim \sqrt{V(\phi(z))}~.
\eeqa
So, the characteristic length (thickness) for varying $\phi(z)$ from two adjacent minima $-v/2$ to $v/2$ in NMSSM can be estimated to be
\beqa
\delta^{-1} \sim \f{\sqrt{V}}{v}\sim \f{\sqrt{\ka A_\ka s^3}}{s}\sim A_\ka~,
\eeqa
while the energy density of the DWs is
\beqa
\rho=\[\phi^\pr(z)\]^2\sim V(z)\sim \ka A_k s^3~.
\eeqa
 The tension of DW can be estimated in terms of the height of the potential energy $V_0$
separating the degenerate minima and the thickness $\delta$ by the relation $\sigma\sim \delta V_0$, giving
$\sigma\sim (\ka A_k s^3)/A_\ka \sim A_\ka^3/\ka^2$. Such an estimation can be helpful to understand some of the subsequent numerical results.

When the exponentially damped friction force of DWs becomes irrelevant after the temperature of the background radiations becomes less than the mass of particles that interact with DWs, the dynamics of DWs is dominated by the tension force, which stretches them up to the horizon size. The evolution of DWs in this regime can be described by the scaling solution, which states that their energy density evolves according to the simple scaling law and there is about one DW per Hubble horizon.
The energy density of DWs in the scaling regime decreases much more slowly than that of cold matters and radiations so that they gradually dominate the energy density of the Universe, drastically altering the subsequent evolution of the Universe.

\section{\label{sec-3} Collapse of DWs from explicit $Z_3$ breaking}

 The formation of DWs from spontaneously discrete $Z_3$ breaking can be problematic in cosmology if they persist till now. They need to collapse before primordial Big Bang nucleosynthesis~(BBN). If the discrete symmetry is broken explicitly by $1/M_P$ suppressed interactions, such operators can lead to a divergent two- or three-loop diagrams that contribute a term linear in $S$, destabilizing the Planck/weak hierarchy unless the coupling is tiny. However, such tiny couplings conflict with the previous constraints from nucleosynthesis~\cite{hep-ph:9604428}.
  Therefore, explicitly $Z_3$ breaking model should have nonvanishing $\mu_0\neq 0$ or $\mu^\pr\neq 0$ with $\ka\neq0$ in the low-energy effective superpotential
  \beqa
W_{\not{Z}_3}\supseteq \mu_0 H_u H_d+ \f{1}{2} \mu^\pr S^2~,
\eeqa
and, at the same time, the absence of the $1/M_P$ suppressed $S^4/M_P$ operator~\footnote{If a $Z_2$ R-symmetry is imposed on the non-renormalizable operators~\cite{hep-ph:9809475}, the stability and the DW problems can be solved together. }. Although the superpotential will not introduce new terms by the nonrenormalization theorem,  such terms and soft SUSY breaking terms
\beqa
-{\cal L}_\mathrm{soft;NMSSM} &\supseteq & -{\cal L}_\mathrm{soft;MSSM}
+ m_{S}^2 | S |^2+\lambda A_\lambda\, H_u \cdot H_d\; S + \frac{1}{3} \kappa A_\kappa\,
S^3 \nn\\
&+&\( B\mu_0 \, H_u \cdot H_d + \frac{1}{2}B'\mu'\, S^2
+ \mathrm{h.c.} \)\; ,
\eeqa
always appear in the low-energy effective action for the light superfields~\cite{ellwanger:0803.2962} after integrating out the messenger/sequestered sector for the SUSY breaking. However, such $\mu_0$ and $\mu^\pr$ terms can again lead to
the existence of tadpole diagrams for $S$, possibly causing the shift in the potential for $S$ to slide to large values far above the weak scale. In fact, a more general conclusion in~\cite{hep-ph:9604428} states that, to ensure a model with singlets will be natural, typical symmetry (such as the gauged R-symmetry or modular symmetry for target space duality~\cite{hep-ph:9506359} instead of ordinary gauge symmetry) is needed to forbid odd-dimension terms in the Kahler potential $K$ and even-dimension terms in the superpotential $\hat{W}$, while any extra odd-dimension operators in $\hat{W}$ or even-dimension operators in $K$ are not harmful to the gauge hierarchy. So, the low-energy effective $\mu^\pr S^2$ and $\mu_0 H_u H_d$ operators should be absent in the original superpotential $\hat{W}$ for supergravity to avoid problematic singlet tadpole problems and they can only come from the Kahler potential. Such small $\mu_0$ and $\mu^\pr$ terms can be the natural consequence of small geometrical breaking of larger local superconformal symmetry from the bilinear dimensionless holomorphic terms in the frame function for the canonical superconformal supergravity (CSS) model~\cite{1004.0712}, when the scale-invariant superpotential is embedded into it.

\subsection{Explicitly $Z_3$ violation term from geometrical superconformal breaking}
The form of Poincare supergravity can be obtained from the underlying superconformal theory, which in addition to local supersymmetry of a Poincare supergravity-has also extra local symmetries: Weyl symmetry, $U(1)_R$ symmetry, special conformal symmetry and special supersymmetry. General theory of supergravity in an arbitrary Jordan frame was derived in ~\cite{SUGRA:SC} from the $SU(2,2|1)$ superconformal theory after gauge fixing. It was noted in~\cite{1004.0712} that the embedding of the globally superconformal theory into supergravity in the Jordan frame can be realized by simply adding the action of the global SUSY, interacting with gravity with conformal scalar-curvature coupling, to the action of supergravity.
The  scalar-gravity part of the ${\cal N}=1$, $d=4$ supergravity in a generic Jordan frame with frame function $\Phi(z, \bar z)$ and superpotential $W(z)$ is given by ~\cite{1004.0712}
\begin{eqnarray}
\mathcal{L}_{J}^{\rm s-g} = \sqrt{-{g}_{J}}\left[\Phi \left(  -\frac{1}{6} {R}({g}_J)+ {\cal A}^2_\mu(z, \bar z) \right )
+\left( \frac{1}{3}\Phi g_{\alpha\bar\beta }-\frac{\Phi
_\alpha \Phi _{\bar\beta }}{\Phi }\right) \hat {\partial}_\mu z^\alpha \hat {\partial}^\mu \bar z^{\bar\beta }
-V_J \right],
  \label{Turin-4}
\end{eqnarray}
with
\begin{eqnarray}
&& \Phi_\alpha \equiv {\frac{\partial }{\partial z^\alpha }}\Phi(z, \bar z),\qquad  \Phi_{\bar\beta } \equiv {\partial\over \partial{\bar z}^{\bar\beta }}\Phi(z, \bar z), \quad  g_{\alpha \bar \beta}= {\frac{\partial^2  \mathcal{K}(z, \bar z)}{\partial z^\alpha  \partial \bar z^{\bar \beta}}}\equiv \mathcal{K}_{\alpha \bar \beta} (z, \bar z),
\end{eqnarray}
and $\mathcal{A}_\mu$ depends on scalar fields as follows:
\be
{\cal A}_\mu(z,{\bar z}) \equiv
-\frac{i}{2\Phi }\,
  \left( \hat {\partial}_\mu z^\alpha\partial_\alpha\Phi
  - \hat {\partial} _\mu \bar z^{\bar\alpha }\partial_{\bar\alpha }\Phi \right).
 \label{A} \ee

 For a particular choice of the frame function, the kinetic scalar terms are canonical when the on-shell auxiliary axial-vector field ${\cal A}_\mu$ vanishes. This requires that
\begin{equation}
 \Phi(z,{\bar z}) = -3 e^{-(1/3) \mathcal{K}(z,{\bar z})} = -3 + \delta_{\alpha\bar\beta }z^{\alpha} \bar z^{\bar\beta } +J(z) +\bar J({\bar z})\,  \ .
\label{Phi}
\end{equation}
These $J\left(z\right) +\overline{J}\left( \overline{z}\right) $ terms
in the frame function not only break the
continuous ${R}$ symmetry, but also break the discrete
${Z}_{3}$ symmetry.

In addition to the pure supergravity part in the total supergravity action, which breaks superconformal symmetry, the part of the action describing chiral
and vector multiplets coupled to supergravity has a much larger local superconformal symmetry. Such a superconformal invariance needs to be broken to make most of the particles massive.
However, this symmetry is broken down to the local Poincare supersymmetry only by the part of the action
describing the self-interacting supergravity multiplet, sometimes requiring additional superconformal symmetry breaking sources in the matter sector. It was proposed in \cite{1008.2942} that the superconformal symmetry of the matter multiplets in the supergravity action can be broken geometrically without introducing dimensional parameters into the underlying superconformal action, in which the moduli space of chiral fields including the compensator field is no longer flat.

We adopt the setting of Jordan frame NMSSM with the following real function
\beqa
{\cal N}=-|X^0|^2+|H_u|^2+|H_d|^2+|S|^2+\chi\( \f{\bar{X}_0}{X_0}S^2+h.c.\)~,
\eeqa
that describes the Kahler manifold of the embedding space, which leads to the superpotential in Jordan frame
\be
V_J=G^{\alpha \bar \beta}W_\alpha  \bar W_{\bar \beta} + \f{1}{2}({\rm Re} f)^{-1\,AB} {\cal P}_A{\cal P}_B \ ,
\label{VJ}
\ee
with $G^{\alpha \bar \beta}$ the matter part of the inverse  metric $G^{I\bar J}$ of the enlarged space including the compensator. The general expression of $G^{\alpha \bar \beta}$ can be calculated to be~\cite{1008.2942}
\beqa
G^{\alpha \bar{\beta }} &=&\delta ^{\alpha \bar{\beta }}-\frac{%
4\chi ^{2}\delta ^{\alpha \bar{\lambda }}\delta ^{\sigma \bar{%
\beta }}a_{\sigma \zeta }\bar{a}_{\bar{\lambda }\bar{\xi }%
}z^{\zeta }\bar{z}^{\bar{\xi }}}{\left[ 3M_{P}^{2}-\chi \left(
a_{\gamma \eta }z^{\gamma }z^{\eta }+\bar{a}_{\bar{\gamma }%
\bar{\eta }}\bar{z}^{\bar{\gamma }}\bar{z}^{\bar{%
\eta }}\right) +4\chi ^{2}\delta ^{\gamma \bar{\eta }}a_{\gamma \theta }%
\bar{a}_{\bar{\eta }\bar{\rho }}z^{\theta }\bar{z}^{%
\bar{\rho }}\right] }\,
\eeqa
for the real function
\begin{equation}
\mathcal{N}\left( X,\bar{X}\right) =-\left| X^{0}\right| ^{2}+\left|
X^{\alpha }\right| ^{2}-\chi \left( a_{\alpha \beta }\frac{X^{\alpha
}X^{\beta }\bar{X}^{\bar{0}}}{X^{0}}+\bar{a}_{\bar{%
\alpha }\bar{\beta }}\frac{\bar{X}^{\bar{\alpha }}\bar{X}%
^{\bar{\beta }}X^{0}}{\bar{X}^{\bar{0}}}\right)\,.
\end{equation}
So, for our case, we have
\beqa
G^{S \bar{S}}&=&\delta^{S \bar{S}}-\f{4\chi^2 \bar{S} S}{\left[ 3M_{P}^{2}-\chi\(S^2+\bar{S}^2\)+4\chi^2\( \bar{S} S\)\right]}\nn\\
&\approx& 1-\f{4\chi^2  |S|^2}{3M_P^2}-\f{\chi(S^2+S^{*2})}{9 M_P^4}+\f{16 \chi^4|S|^4}{9 M_P^4}+\cdots~.
\eeqa
The SUSY preserving part of the scalar potential receives an additional term
\beqa
\Delta V_J\supseteq -\[\f{4\chi^2  |S|^2}{3M_P^2}+\f{4\chi^3|S|^2(S^2+S^{*2})}{9 M_P^4}-\f{16 \chi^4|S|^4}{9 M_P^4}+\cdots\]\left|\lambda \left(H_u^+ H_d^--H_u^0
H_d^0\right) + \kappa S^2 \right|^2.\nn\\
\label{Z3:violationSS}
\eeqa
We can see that the leading $Z_3$ breaking term in this scalar potential that proportional to $(S^2+S^{*2})$ is suppressed by $1/M_P^4$. Similarly, for real function ${\cal N}(X,\bar{X})$ of the form
\beqa
{\cal N}(X,\bar{X})=-|X^0|^2+|H_u|^2+|H_d|^2+|S|^2+\chi\( \f{\bar{X}_0}{X_0}H_u H_d+h.c.\)~,
\eeqa
in the Jordan frame for NMSSM, we have
\beqa
G^{H_u \bar{H}_u}&=&\delta^{H_u \bar{H}_u}-\f{\chi^2 \bar{H}_u H_u}{\left[ 3M_{P}^{2}-\chi\(H_u H_d+\bar{H}_u\bar{H}_d\)+\chi^2\( \bar{H}_u H_u+\bar{H}_d H_d\)\right]}~,\nn\\
G^{H_d \bar{H}_d}&=&\delta^{H_d \bar{H}_d}-\f{\chi^2 \bar{H}_d H_d}{\left[ 3M_{P}^{2}-\chi\(H_u H_d+\bar{H}_u\bar{H}_d\)+\chi^2\( \bar{H}_u H_u+\bar{H}_d H_d\)\right]}~,\nn\\
G^{H_u \bar{H}_d}&=&-\f{\chi^2 \bar{H}_d H_u}{\left[ 3M_{P}^{2}-\chi\(H_u H_d+\bar{H}_u\bar{H}_d\)+\chi^2\( \bar{H}_u H_u+\bar{H}_d H_d\)\right]}~,\nn\\
G^{H_d \bar{H}_u}&=&-\f{\chi^2 \bar{H}_u H_d}{\left[ 3M_{P}^{2}-\chi\(H_u H_d+\bar{H}_u\bar{H}_d\)+\chi^2\( \bar{H}_u H_u+\bar{H}_d H_d\)\right]}~.
\eeqa
So, the scalar potential for NMSSM is given by
\beqa
V_J&=&\la^2 |S|^2 |H_d|^2 G^{H_u \bar{H}_u}+ \la^2 |S|^2 |H_u|^2 G^{H_d \bar{H}_d}+ \la^2 |S|^2 G^{H_d \bar{H}_u} \bar{H}_d H_u+\la^2 |S|^2G^{H_d \bar{H}_u} \bar{H}_d H_u\nn\\
&+&|\la H_u H_d+\ka S^2|^2+\f{1}{8}(g_1^2+g_2^2)(|H_d|^2-|H_u|^2)^2+\f{1}{2}{g_2^2}|H_u H_d^\da|^2~,
\eeqa
with the correction
\beqa
\Delta V_J&\supseteq& -4\[\f{\chi^2 }{3M_{P}^{2}}+\f{\chi^3 \(H_u H_d+\bar{H}_u\bar{H}_d\)}{9M_{P}^{4}}+\f{\chi^4 \(|{H}_u|^2 +|{H}_d|^2\)}{9M_{P}^{4}}\]\la^2 |S|^2 |H_d|^2|H_u|^2~.\nn\\
\label{Z3:violationHuHd}
\eeqa
Therefore, the leading $Z_3$ breaking term that proportional to $(H_u H_d+\bar{H}_u\bar{H}_d)$ in the scalar potential can be seen to be suppressed by $1/M_P^4$.

The frame function $\Phi(z^a,\bar{z}^a)$ and consequently the Kahler potential can relate to ${\cal N}(X,\bar{X})$ by the gauge fixing
\beqa
X^0=\bar{X}^{\bar{0}}=\sqrt{3}M_P,~~~X^a=z^a~,
\eeqa
with ${\cal N}(X,\bar{X})=\Phi(z^a,\bar{z}^a)=-3e^{-K(z,\bar{z})/3}$.

In many popular SUSY breaking mechanisms, the total superpotential can be divided into the hidden sector and visible sector with $W=W_{vis}+W_{hid}$ for $\langle W_{vis}\rangle \ll \langle W_{hid}\rangle$, as the VEV is obviously dominated by the fields responsible for SUSY breaking even though all fields in the theory are summed over. So, under Kahler transformation to render the new Kahler potential canonical
\beqa
K(z,\bar{z})\ra K(z,\bar{z})+J(z)+\bar{J}(\bar{z}), ~~W(z)\ra  e^{-J(z)/M_P^2} W(z)~,
\eeqa
the superpotential will receive a correction of the from
\beqa
\delta W&\simeq& \f{J(z)}{M_P^2} W\simeq\f{J(z)}{M_P^2}\langle W_{hid}\rangle
\simeq m_{3/2} J(z)~.
\eeqa
Here the gravitino mass $m_{3/2}$ is
\beqa
m_{3/2}=e^{\f{K}{2M_P^2}}\f{\langle W\rangle}{M_P^2}\simeq \f{\langle W\rangle}{M_P^2}~.
\eeqa

For $J(z)=\chi S^2$ (or $\chi H_u H_d$), the superpotential will receive (non-Planck scale suppressed) $Z_3$ breaking terms $\chi m_{3/2}S^2$ (or $\chi m_{3/2} H_u H_d$), respectively. Note that unpreferable tadpole term $\xi_F S $ will not be present in such $Z_3$ explicitly breaking superpotential because no dimensional parameters are introduced in ${\cal N}(X,\bar{X})$, which can be seen as an advantage of this CSS approach.

After the Kahler transformation with new frame function $\Phi^\pr(z,\bar{z})$, the action in the Jordan frame is given by
\beqa
\f{{\cal L}_J}{\sqrt{-g}}=-\f{1}{6}\Phi^\pr(z,\bar{z})R(g_J)-\delta_{\al\bar{\beta}}\pa_\mu z^\al \pa^\mu z^{\bar{\beta}}-V_J^\pr(z,\bar{z})~,
\eeqa
with $V_J^\pr$ the scalar potential for the superpotential $e^{-J} W$. It is obvious that the matter part of supergravity action contains nonminimal gravitational couplings of scalars to the Ricci scalar curvature in additional to the bosonic part of the supergravity action. To absorb the nonminimal gravitational couplings, we need to rescale the metric. Under the transformation
\beqa
\tl{g}_{\mu\nu}=e^{2\sigma(x)} {g}_{\mu\nu}~,
\eeqa
the Ricci scalar changes as
\beqa
\tl{R}=e^{-2\sigma(x)} \[R-2(d-1)\nabla^2(\sigma)-(d-1)(d-2)(\nabla\sigma)^2\]~,
\eeqa
with $d=4$, $\nabla^2\sigma=g^{\mu\nu}\nabla_\mu\nabla_\nu\sigma$ and $(\nabla\sigma)^2=g^{\mu\nu}(\nabla_\mu\sigma)(\nabla_\nu\sigma)$. So, the metric can be rescaled as followings
\beqa
e^{2\sigma}\equiv \Omega^2=1-\f{1}{3 M_P^2}\left\{|S|^2 + |H_u|^2 + |H_d|^2\right\}~,
\eeqa
to change the Jordan frame into the Einstein frame~\footnote{We should note that, without the Kahler transformation, the field variables in the Einstein frame with the corresponding metric rescaling involving $J(z)$ are very complicate. }. After metric rescaling, the scalar potential in the Einstein frame is given by
\beqa
V_E=\f{V_J^\pr}{\Omega^4}\simeq \[1+\f{2}{3 M_P^2}\left\{|S|^2 + |H_u|^2 + |H_d|^2\right\}\]V_J^\pr(e^{-J} W),
\eeqa
with $V_J^\pr(e^{-J} W)$ the SUSY scalar potential for the superpotential $e^{-J} W$. General discussions for nonminimal gravitational couplings involving the complex scalars can be found in our previous work~\cite{WW:2104.04682}, in which the polar variables ${R},{\Theta}$ can be used to simplify the transformations.
As anticipated, in small field values $\phi\ll M_P$, the new variables in the Einstein frame can reduce to the original field variables. The $m_{3/2} J(z)$ term in the superpotential can give the dominant (non-Planck scale suppressed) $Z_3$ breaking contribution in the Einstein frame scalar potential in comparison to the $Z_3$ breaking terms in~(\ref{Z3:violationSS}) and~(\ref{Z3:violationHuHd}).

\subsection{Additional explicitly $Z_3$ breaking terms after SUSY breaking}
In the CSS approach with geometrical superconformal breaking in the matter sector, no tadpole terms in the superpotential and scalar potential will be generated after the Kahler transformation with SUSY breaking.  The mediation mechanism of SUSY breaking that is responsible for the soft SUSY breaking parameters of NMSSM fields should also not reintroduce the tadpole terms. Unlike some gravity mediation mechanism, which can potentially trigger the generation of tadpole terms again, the tadpole terms in gauge mediated SUSY breaking ~(GMSB) mechanism can be forbidden by some discrete symmetry for the messenger sector. For example, the economical realization~\cite{NMSSM:YM1,NMSSM:YM2} (without the introduction of additional gauge fields) can be adopted
\beqa
W\supseteq \sum\limits_{i=0}^n \la_{X;i} X(\bar{\Phi}_{2i} {\Phi}_{2i}+\bar{\Phi}_{2i+1}{\Phi}_{2i+1})+\sum\limits_{i=0}^n\la_{S;i} S \bar{\Phi}_{2i}{\Phi}_{2i+1}~,
\eeqa
to avoid the tadpole problem, which can preserve the $Z_3$ discrete symmetry in the messenger sector by introducing double families of messengers (in ${\bf 5\oplus \bar{5}}$ representation of $SU(5)$). In this case, the general soft SUSY breaking parameters in the NMSSM, such as $B$ and $B^\pr$, can be estimated to leading order as
\beqa
B^\pr\simeq \f{1}{16\pi^2} \f{F_X}{M_{m}} 5\sum\limits_{j=1}^N\la_{S;j}^2,~~~B=0~,
\eeqa
for the VEV of spurion $\langle X\rangle=M_m+\theta^2 F_X$. For a general messenger sector with proper messenger-matter interactions involving the $H_u$ or $H_d$ fields, the $B$ parameter will receive similarly contribution of order the soft SUSY breaking masses $m_{soft}\sim \f{1}{16\pi^2} \f{F_X}{M_{m}}\sim {\cal O}({\rm TeV})$.

In our subsequent studies, we will not specify concretely which SUSY breaking mechanism is used. The values $B\simeq B^\pr\sim m_{soft}$ with $m_{soft}=\sqrt{m_{\tl{t}_1}m_{\tl{t}_2}}$ are adopted in our numerical results for such explicitly $Z_3$ breaking soft SUSY parameters.

\section{\label{sec-4} Collapse of DWs and GWs}

The $\chi m_{3/2} S^2$ related terms in the scalar potential break explicitly the $Z_3$ symmetry. The scalar potential relevant to such a term is given by
\beqa
\Delta {\cal L}_{S^2}&\supseteq& \[|\mu^\pr S|^2+\left\{\f{}{}\mu^\pr S^*\(\lambda H_u H_d + \kappa S^2\)+\frac{B'\mu'}{2}\, S^2 +h.c.\right\}\]\[1+\f{2\left\{|S|^2 + |H_u|^2 + |H_d|^2\right\}}{3 M_P^2}\],\nn\\
&-&\[\f{4\chi^2  |S|^2}{3M_P^2}+\f{4\chi^3|S|^2(S^2+S^{*2})}{9 M_P^4}-\f{16 \chi^4|S|^4}{9 M_P^4}+\cdots\]\left|\lambda H_u H_d + \kappa S^2\right|^2,\nn\\
&-&\f{8 \chi^2  |S|^2}{9 M_P^4}\left\{|S|^2 + |H_u|^2 + |H_d|^2\right\}+{\cal O}(1/M_P^6)~,
\eeqa
with $\mu^\pr=2\chi m_{3/2}$. We will neglect the subleading $M_P$ suppressed $Z_3$ breaking terms in our subsequent studies for the collapsing of DWs. The vacuum energy differences between $v_i$ and $v_j$ are defined as
\beqa
(V_{bias})_{i,j} = V_{v_i}- V_{v_j}~, ~~{\rm for} ~~(i,j=1,\omega,\omega^2)
\eeqa
which satisfies $(V_{bias})_{1,\omega^2}-(V_{bias})_{1,\omega}=(V_{bias})_{\omega,\omega^2}$.

We can calculate the energy difference between adjacent vacuum
\beqa
\Delta_{1,\omega} V_{\not{Z}_3; S^2}&=&\Delta_{1,\omega^2} V_{\not{Z}_3; S^2}\nn\\
&=&\[2\mu^\pr s \(\ka s^2-\la v_u v_d\)+\mu^\pr m_{soft} s^2 \](1-\cos\f{2\pi}{3})\nn\\
&=&3 \mu^\pr s\(\ka s^2-\la v_u v_d\)+\f{3}{2}\mu^\pr m_{soft} s^2~,\nn\\
\Delta_{\omega,\omega^2} V_{\not{Z}_3; S^2}&=& 0~.
\eeqa
Requiring the $v_1$ vacua to be the true minimum leads to the constraints
$\Delta_{1,\omega} V_{\not{Z}_3; S^2}<0$ and $\Delta_{1,\omega^2} V_{\not{Z}_3; S^2}<0$, which can be satisfied with
\beqa
\mu^\pr<0~,~\qquad{\rm or}~\quad\quad \(\ka s^2-\la v_u v_d\)+\f{1}{2} m_{soft} s<0~.
\label{vacuum:order}
\eeqa
The degeneracy of the two minimum is the consequence of CP conjugation of the $\omega$ and $\omega^2$ vacua.  When the conditions in Eq.(\ref{vacuum:order}) are satisfied, we have the order of the vacuum energies $V_{v_{\omega^2}} =V_{v_{\omega}}>V_{v_1}$ with ${v_1}$ the true vacua and
\beqas
|(V_{bias})_{1,\omega^2}|=|(V_{bias})_{1,\omega}|,~~~|(V_{bias})_{\omega,\omega^2}|=0, \eeqas
 for the biases.

With bias terms, the energy difference between the neighboring vacua acts as a volume pressure $p_V\sim V_{bias}$ on the wall, which tends to shrink the higher-energy domains until the whole space is filled by the only true vacuum, eventually collapsing the wall network. The
collapse of DWs happens when this pressure force becomes greater than the tension force $p_T\sim \sigma/R_{wall}$. In our case with two degenerate false vacua, the volume pressures from $(V_{bias})_{1,{\omega}^2}$ and $(V_{bias})_{1,{\omega}}$ are of the same strength, and
they lead the true $v_1$ vacuum to expand against the false $v_{{\omega}}$ and $v_{{\omega}^2}$
vacua regions, reducing at the same rate the occupancy of such false
vacuum states and leading to the fragmentation of the wall network, which will collapse.

If the conditions in Eq.(\ref{vacuum:order}) are not satisfied, the potential has two degenerate true minima $v_{\omega}$ and $v_{\omega^2}$ .
 To spoil the problematic degeneracy between the $\omega$ and $\omega^2$ vacua, one can introduce additional messengers $\psi^\pr$, $\bar{\psi}^\pr$, which are charged under some hidden gauge group $SU(N)_h$ and $Z_3$. The $Z_3$ discrete symmetry can be anomalous under $SU(N)_h$, leading to additional contributions to bias of order $\Delta_H V\sim \lambda_H^4$~\cite{Hamaguchi:2011nm} so as that the degeneracy between the $\omega$ and $\omega^2$ vacua is spoiled (assuming that $V_{v_{\omega^2}}\gtrsim V_{v_{\omega}}$). Then, the vacuum energies and the biases satisfy
\beqa
&&V_{v_1}>V_{v_{\omega^2}}\gtrsim V_{v_\omega},\nn\\&&(V_{bias})_{1,\omega}|\gtrsim |(V_{bias})_{1,\omega^2}|\gg |(V_{bias})_{\omega,\omega^2}|,
\label{DW:01-2}
\eeqa
In such a scenario, due to the fact that the true vacuum $v_{\omega}$ is nearly degenerate with the false vacuum $v_{\omega^2}$, the annihilations driven by $(V_{bias})_{1,{\omega}^2}$ and $(V_{bias})_{1,{\omega}}$ are of similar strengths, increasing almost identically the $v_\omega$ and $v_{\omega^2}$ regions by volume pressure to diminish the original false vacuum $v_1$ region. As $v_{\omega^2}$ is still a false vacuum, the tiny $(V_{bias})_{\omega,\omega^2}$ driven volume pressure on $v_{\omega^2}$ can slowly push the walls to increase the true vacuum $v_{\omega}$ region. So, after the period with quickly increment of the $v_{\omega^2}$ region driven by the $(V_{bias})_{1,{\omega^2}}$ from original $v_1$ region, the $v_{\omega^2}$ region tends to diminish slowly by
the tiny $(V_{bias})_{\omega,\omega^2}$ driven volume pressure. In this case the decay time for DWs is dominated by the slow $(V_{bias})_{\omega,\omega^2}$ driven collapse. Collapse of DWs in such a scenario with vacuum structure satisfying~(\ref{DW:01-2}) is sensitive to the small contributions $(V_{bias})_{{\omega},{\omega}^2}$ that lift the degeneracy between $v_\omega$ and $v_{\omega^2}$. Not specifying the details of the degeneracy lifting mechanism, we will not discuss such a scenario in this work and only concentrate on the scenario with $v_1$ the true vacuum that was discussed in previous paragraphs.

The collapse of DWs happens when the volume pressure force becomes greater than the tension force.
Assuming  that DWs have reached the scaling regime, we can estimate the decay time
\beqa
t\approx C_{ann}{\cal A}\f{\sigma}{\epsilon}\simeq C_{ann}{\cal A}\f{\sigma}{|\Delta V_{\not{Z}_3; S^2}|}~,
\label{DW:LT}
\eeqa
with ${\cal A}\sim 1.2$ for $N=3$ and $C_{ann}$ a coefficient of ${\cal O}(1)$ that takes value $C_{ann}=
5.02\pm 0.44$, based on the simulation of axion models with $N=3$. We can estimate the decay time as $t\sim C_0 (\chi m_{3/2})^{-1}$ by the following observations:  $\sigma\sim m_{soft}^3$ while $V_{bias}\sim C_1 (\chi m_{3/2}) m_{soft}^3$ with $C_0,C_1$ some constant coefficients. So, the decay time can be dominantly determined by the value of $\chi m_{3/2}$.

It is well known that DWs cannot be formed from the beginning if the bias terms are sufficiently large.
The prediction of percolation theory gives a necessary condition for the formation of large scale DWs with the presence of bias term~\cite{percolation}
\beqa
\f{V_{bias}}{V_0}<\ln\(\f{1-p_c}{p_c}\)=0.795~.
\label{percolation}
\eeqa
We check that this constraint is safely satisfied for the NMSSM scalar potential with  small bias terms, for example, for $\mu^\pr S^2/2$ case with the choice $\chi m_{3/2}\sim 10^{-12}$ {\rm eV} adopted in our numerical studies.

Similarly, the $\mu_0\equiv\chi m_{3/2} H_u H_d$ term that explicitly violates the $Z_3$ discrete symmetry will also cause the collapse of DWs. (We have $\mu_{eff}=\la s+\mu_0\approx \la s$ as the term $\la s$ gives the dominant contribution to $\mu_{eff}$.) The scalar potential relevant to such a term is given by
\beqa
\Delta {\cal L}_{H_u H_d}&\supseteq& \left\{\[|\mu_0|^2+\mu_0 \lambda (S+S^*)\]\left(\left|H_u\right|^2 +\left|H_d\right|^2\right)+(B\mu_0 \, H_u \cdot H_d+h.c.)\right\}\nn\\
&&\[1+\f{1}{3 M_P^2}2\left\{|S|^2 + |H_u|^2 + |H_d|^2\right\}\],\nn
\\
&-& 4\[\f{\chi^2 }{3M_{P}^{2}}+\f{\chi^3 \(H_u H_d+\bar{H}_u\bar{H}_d\)}{9M_{P}^{4}}+\f{\chi^4 \(|{H}_u|^2 +|{H}_d|^2\)}{9M_{P}^{4}}\]\la^2 |S|^2 |H_d|^2|H_u|^2,\nn\\
&-& 8\f{\chi^2 }{9M_{P}^{4}}\la^2 |S|^2 |H_d|^2|H_u|^2~.
\eeqa
We can calculate the energy difference between adjacent vacuum
\beqa
\Delta_{1,\omega} V_{\not{Z}_3; H_u H_d}&=&\Delta_{1,\omega^2} V_{\not{Z}_3; H_u H_d}\nn\\
 &=& 2\[\lambda \mu_0 v^2 s+m_{soft}\mu_0v_uv_d\](1-\cos\f{2\pi}{3})\nn\\
&\simeq &3 \lambda \mu_0 s v^2+3m_{soft}\mu_0\f{v^2}{\tan\beta}\approx 3 \mu_0v^2 \( \mu_{eff} +\f{m_{soft}}{\tan\beta}\),\nn\\
\Delta_{\omega,\omega^2} V_{\not{Z}_3; S^2}&=& 0~.
\eeqa
From the electroweak symmetry breaking condition in~Eq.(\ref{EWSB}), we have $\mu_{eff}\gg \f{m_{soft}}{\tan\beta}$ so that the contribution to $\Delta_{1,\omega;\omega^2} V$ from soft SUSY breaking $B$ term is always negligible. Requiring the $v_1$ vacua to be the true minimum leads to the constraints $\mu_0<0$.
Again, in the scaling regime, the decay time of DWs can be estimated as
\beqa
t\approx C_{ann}{\cal A}\f{\sigma}{\epsilon}\sim C_{ann}{\cal A}\f{\sigma}{|\Delta V_{\not{Z}_3; H_u H_d}|}~.
\eeqa
Similarly, the decay time can be dominantly determined by the value of $\chi m_{3/2}$. The constraint~(\ref{percolation}) is also safely satisfied for the NMSSM scalar potential with tiny explicit $Z_3$ breaking coefficient $\chi m_{3/2}\sim 10^{-12}$ {\rm eV} for $\mu_0 H_u H_d$ case.

The collapse of DWs should occur before they overclose the Universe  and their decay products  should at the same time not destroy light elements created at the epoch of BBN. So,  we have the constraints
\beqa
t\lesssim \min(0.01~{\rm sec}, t_{dom})=\min(1.4\tm 10^{13}~{\rm eV^{-1}}, t_{dom})~,
\label{constraints:DWt}
\eeqa
with
\beqa
t_{dom}=\f{3 M_P^2}{4{\cal A}\sigma}\simeq  2.93 \tm 10^{3}{\rm  sec}{\cal A}^{-1}\(\f{\sigma}{\rm TeV^3}\)^{-1}\simeq 0.45\tm 10^{19} {\rm eV^{-1}}{\cal A}^{-1}\(\f{\sigma}{\rm TeV^3}\)^{-1}.
\eeqa
The BBN constraint is always more stringent than that of $t_{dom}$ unless $\sigma\gtrsim 10^6 {\rm TeV^3}$.
\subsection{GWs from the collapse of DWs}
The collisions of the DWs can deviate from the spherical symmetry, so the energy stored inside DWs can not only be released into light degrees of freedom but also into GWs when DWs collapse. The intensity of the GWs decreases as the Universe expands since the amplitude is redshifted for each mode. So, one may use a dimensionless quantity $\Omega_{gw}(f)$ to characterize the intensity of GWs.
The density parameter $\Omega_{gw}(f)$ of the GWs can be defined as the fraction between the GWs energy density per logarithmic frequency interval and the total energy density of the Universe
\beqa
\Omega_{gw}(f)\equiv\f{1}{\rho_c}\f{d \rho_{gw}}{d\log f}~,
\eeqa
with $\rho_{gw}$ as the energy density of the GWs, $\rho_c$ the critical energy density and $f$ the
frequency of GWs. As the GWs are predominantly sourced by horizon-scale structures in the DW network, the GWs emitted are expected to peak at the frequency corresponding to the horizon scale at the decay time.

Due to the redshift by the cosmic expansion, the present density parameter $\Omega_{gw}^0$  and the frequency of the GWs $f_0$ can be obtained from the quantities at the formation of the GWs (denoted with $'*'$) as
\beqa
\Omega_{gw}^0=\Omega_{gw}^*\(\f{a_*}{a_0}\)^4\(\f{H_*}{H_0}\)^2~,~~f_0=f_*\(\f{a_*}{a_0}\).
\eeqa

The peak frequency today can be estimated by
\beqa
f_{peak}&=&\f{a(t_*)}{a(t_0)} H_*\f{f_*}{H_*}~,\nn\\
&=& 1.14 nHz \(\f{10.75}{g_*,s}\)^{1/3}\(\f{g_*}{10.75}\)^{1/2}\(\f{10^{13}{\rm eV}^{-1}}{t}\)^{1/2}~,
\eeqa
with $g_{*}$ and $g_{*s}$ the effective relativistic degrees of freedom for the energy density and the entropy density, respectively. Here we use the following formula to transform the temperature into the Hubble parameter
\beqa
H=\f{1}{2t}=\sqrt{\f{8\pi^3}{90}g_*}\f{T^2}{M_P}=1.66 {g_*}^{\f{1}{2}}\f{T^2}{M_P}~,
\eeqa
in the radiation dominant era with $M_P=1.22\tm 10^{19} {\rm GeV}$.

For $s,m_{soft}\sim {\cal O}(1 )$ TeV, $\la,\ka\sim {\cal O}(1)$ and $\tan\beta\sim {\cal O}(10)$, the value of $\chi m_{3/2}$ should be of order $10^{-12}$ eV for $\sigma\sim {\cal O}({\rm TeV })^3$. Given that the gravitino mass $m_{3/2}$ cannot be much lighter than ${\cal O}(1)$ eV, the $\chi$ parameter should be much smaller than unity. Such tiny value can be the consequence of higher-dimensional operators in the Kahler potential with $U(1)_R$ symmetry. For example, assuming $R(\Phi)=16/3$ and $R(\Phi^\pr)=4$, the Kahler potential adopted in~\cite{hep-ph:9603301}
\beqa
K\supseteq \Phi^\da\Phi+\Phi^{\pr\da}\Phi^\pr+\(\f{\al}{M_P^2}\Phi\Phi^{\pr\da}H_u H_d+\f{\al^\pr}{M_P^2}\Phi\Phi^{\pr\da} S^2+h.c.\)+\cdots,
\eeqa
can naturally lead to $\chi\sim 10^{-24}$ for $m_{3/2}\sim {\cal O}(1) {\rm GeV}$ and $\al,\al^\pr\sim {\cal O}(1)$ without the tadpole problem, when the R-symmetry breaking scale lies of order $\langle\Phi\rangle\sim\langle\Phi^\pr\rangle\sim 10^{6} {\rm GeV}$. Such Kahler potential can originate from the real function $\mathcal{N}( X,\bar{X})$
\beqa
\mathcal{N}\left( X,\bar{X}\right) =-\left| X^{0}\right| ^{2}+\sum\limits_{\Phi,\Phi^\pr\cdots}\left|X^{\alpha }\right| ^{2}+
\(\al\f{X^\Phi \bar{X}^{\Phi^\pr}}{|X_0^2|} H_u H_d+\al^\pr\f{X^\Phi \bar{X}^{\Phi^\pr}}{|X_0^2|} S^2+h.c.\),
\eeqa
after gauge fixing.

The power of the GWs radiation can be estimated as $\rho_{gw}\sim Pt/t^3\sim G {\cal A}\sigma^2$ by the quadrupole formula, given by $P\sim G \dddot{Q}_{ij} \dddot{Q}_{ij}\sim M^2_{wall}/t^2$ with $M_{wall}=\sigma {\cal A} t^2$ the energy of DWs.  The quadrupole formula cannot be directly applied to DWs, since it is only valid in the far-field regime~\cite{1703.02576}. To obtain the peak amplitude of GWs produced by long-lived DWs, we use the fact that numerical simulations show that the value
\beqa
\tl{\epsilon}_{gw}\equiv \f{1}{G{\cal A}^2\sigma^2}\(\f{d\rho_{gw}}{d\ln f}\)_{peak},
\eeqa
almost keeps the constant value $\tl{\epsilon}_{gw}\simeq 0.7\pm 0.4$~\cite{GW:estimation} after DWs enter into the scaling regime. So, the peak amplitude at the annihilation time of DWs can be estimated as
\beqa
\Omega_{gw}(H_*)_{peak}=\f{1}{\rho_c(H_*)}\(\f{d \rho_{gw}(H_*)}{d\log f}\)_{peak}=\f{8\pi G^2{\cal A}^2\sigma^2\tilde{\epsilon}_{gw}}{2H_*^2}~.
\eeqa
Here the production of GWs is assumed to suddenly terminate at $H=H_*$ and happens during the radiation dominated era. The present density parameter $\Omega_{gw}^0$ can be estimated as~\cite{1703.02576}
\beqa
\Omega_{gw}(t_0)_{\rm peak} h^2&=&\Omega_{rad} h^2\(\f{g_*(T_{ann})}{g_{*0}}\)\(\f{g_{*s0}}{g_{*s}(T_{ann})}\)^{4/3}\Omega_{gw}(t_{ann}),~\nn\\
&\simeq & 3.06 \tm 10^{-14} \(\f{10.75}{g_*}\)^{1/3} \(\f{t}{10^{13}{\rm eV}^{-1}}\)^2\(\f{\sigma}{1 {\rm TeV}^3 }\)^2~,
\label{omega:gw0}
\eeqa
with $\Omega_{rad} h^2= 4.15\tm 10^{-5} $ the density parameter of radiations at the present time. The constraint $t\lesssim t_{dom}$ can set an upper bound for $\Omega_{gw}(t_0)_{\rm peak} h^2$ with
\beqa
\Omega^{upper}_{gw}(t_0)_{\rm peak} h^2\lesssim 10^{-8}~,
\eeqa
for the case with $\sigma\gtrsim 10^6~{\rm TeV^3}$.

The full GW spectrum today can be estimated by
\beqa
\Omega_{gw}(f) h^2=\Omega_{gw}(t_0)_{\rm peak} h^2\left\{\bea{c}\(\f{f}{f_{\rm peak}}\)^3,~~~f<f_{\rm peak}\\ \f{f_{\rm peak}}{f},~~~~~~~f>f_{\rm peak}\eea \right.~~~.
\eeqa

  In previous discussions, it is assumed that the annihilation of DWs happens during the radiation
dominated era. If it happens before reheating, the above estimations can be modified accordingly. Numerical results indicate that such a possibility requires very low reheating temperature of order MeV to explain the recent nHZ GWs background signals. Low reheating temperature of order GeV can be preferred to avoid the overproduction of very light gravitino and the increase of entropy due to the production of moduli fields~\cite{RHTemp}.  However, this low temperature seems to be problematic with baryogenesis, according to which it is commonly assumed that the reheating temperature is at least of the order the electroweak scale ($10^2 {\rm GeV}$)~\cite{LRHT}. Although there are some mechanisms that can explain the baryon asymmetry at very low temperatures, reheating temperature of order MeV still seems not favored.

\subsection{Tension between $nHz$ PTA GW background data and muon $g-2$ anomaly}
 The PTA observations on the frequency of stochastic GW background can set stringent constraints for the collapse time of DWs, which can be mainly determined by the value of $\chi m_{3/2}$~[see the discussions below Eq.(\ref{DW:LT})]. With the very narrow range of the DW collapse time derived from PTA data, the upper bound for the present density parameter $\Omega_{gw}^0$ can set an upper bound for the tension $\sigma$ of DWs by the formula~(\ref{omega:gw0}), which can be further translated into the upper bound for the soft SUSY breaking parameters $m_{soft}$. We can estimate the upper bound of $m_{soft}$ to be $m_{soft}\lesssim 200 {\rm TeV}$. Such an upper bound on $m_{soft}$ can shed light on the UV SUSY breaking mechanisms.

Most importantly, given the BBN upper bounds for the collapse time of DWs, the lower bound for the present density parameter $\Omega_{gw}^0$ can set the lower bounds for $m_{soft}$, which can be estimated to be of order TeV. Such lower bounds can have tension with the SUSY explanation of the muon $g-2$ anomaly because large SUSY contributions to $\Delta a_\mu$ in general need relatively light superpartners.

We know that the~SUSY contributions to muon $g-2$ are dominated by the chargino--sneutrino and the neutralino--smuon loops in MSSM. At the leading order of $\tan\beta$ and $m_W/m_{SUSY}$, they are evaluated as~\cite{Endo:2013bba}
\begin{align}
  \Delta a_{\mu }(\tilde{\mu }_L, \tilde{\mu }_R,\tilde{B})
 &= \frac{\alpha_Y}{4\pi} \frac{m_{\mu }^2 M_1 \mu}{m_{\tilde{\mu }_L}^2 m_{\tilde{\mu }_R}^2}  \tan \beta\cdot
 f_N \left( \frac{m_{\tilde{\mu }_L}^2}{M_1^2}, \frac{m_{\tilde{\mu }_R}^2}{M_1^2}\right). \label{eq:BmuLR} \\
   \Delta a_{\mu }(\tilde{B}, \tilde{H},  \tilde{\mu }_R)
  &= - \frac{\alpha_Y}{4\pi} \frac{m_{\mu }^2}{M_1 \mu} \tan \beta \cdot
  f_N \left( \frac{M_1 ^2}{m_{\tilde{\mu }_R}^2}, \frac{\mu ^2}{m_{\tilde{\mu }_R}^2} \right), \label{eq:BHmuR} \\
  \Delta a_{\mu }(\tilde{B},\tilde{H},  \tilde{\mu }_L)
  &= \frac{\alpha_Y}{8\pi} \frac{m_\mu^2}{M_1 \mu} \tan\beta\cdot
 f_N
 \left( \frac{M_1 ^2}{m_{\tilde{\mu }_L}^2}, \frac{\mu ^2}{m_{\tilde{\mu }_L}^2} \right),
 \label{eq:BHmuL} \\
 \Delta a_{\mu }(\tilde{W}, \tilde{H},  \tilde{\mu}_L)
 &= - \frac{\alpha_2}{8\pi} \frac{m_\mu^2}{M_2 \mu} \tan\beta\cdot
 f_N
 \left( \frac{M_2 ^2}{m_{\tilde{\mu }_L}^2}, \frac{\mu ^2}{m_{\tilde{\mu }_L}^2} \right),
 \label{eq:WHmuL}  \\
 \Delta a_{\mu }(\tilde{W}, \tilde{H}, \tilde{\nu}_\mu)
 &= \frac{\alpha_2}{4\pi} \frac{m_\mu^2}{M_2 \mu} \tan\beta\cdot
f_C
 \left( \frac{M_2 ^2}{m_{\tilde{\nu }}^2}, \frac{\mu ^2}{m_{\tilde{\nu }}^2}  \right) ,
 \label{eq:WHsnu}
\end{align}
 In the previous expressions, we denote $m_\mu$ as the muon mass, $m_{SUSY}$ the SUSY breaking masses and $\mu$ the Higgsino mass, respectively. The~loop functions are defined as
\begin{align}
&f_C(x,y)= xy
\left[
\frac{5-3(x+y)+xy}{(x-1)^2(y-1)^2}
-\frac{2\log x}{(x-y)(x-1)^3}
+\frac{2\log y}{(x-y)(y-1)^3}
\right]\,,
\\
&f_N(x,y)= xy
\left[
\frac{-3+x+y+xy}{(x-1)^2(y-1)^2}
+\frac{2x\log x}{(x-y)(x-1)^3}
-\frac{2y\log y}{(x-y)(y-1)^3}
\right]\,,
\label{moroi3}
\end{align}
which are monochromatically increasing for $x>0,y>0$ with $0\le f_{C,N}(x,y) \le 1$. They satisfy $f_C(1,1)=1/2$ and $f_N(1,1)=1/6$ in the limit of degenerate masses. The~SUSY contributions to the muon $g-2$ will be enhanced for small soft SUSY breaking masses and large value of $\tan\beta$.
It can be seen from the previous formulas that large SUSY contributions to $\Delta a_\mu$ require light sleptons, light electroweakinos and large $\tan\beta$. However, current LHC experiments have already
set stringent constraints on colored sparticles, such as the $2.2~{\rm  TeV}$ bound for gluino and the
$1.4 ~{\rm TeV}$ bound for squarks, making the explanation of the recent muon $g-2$ data fairly nontrivial. SUSY explanations of the muon $g-2$ anomaly can be
seen in the literatures; see~\cite{mg1,mg2,mg3,mg4,mg5,mg6,mg7,mg8,mg9,mg10,mg11,mg12,mg13,mg14,mg15,mg16} and~\cite{mg17,mg18,mg19,mg20,mg21}. The~inclusion of the singlet component in NMSSM will in general give negligible contributions to $\Delta a_\mu$ because of the suppressed coupling of singlino to the MSSM sector. However, the~lightest neutral CP-odd Higgs scalar could give non-negligible contributions to $a_\mu$ if it is quite light~\cite{NMSSM:g-2}. The~positive two-loop contribution is numerically more important for a light CP-odd Higgs at approximately 3~GeV and the sum of both one-loop and two-loop contributions is maximal around $m_{a_1}\sim 6~{\rm GeV}$. However, such a light $a_1$ is constrained stringently by various low-energy experiments, such as CLEO and B-physics.

Given the estimated ${\cal O}({\rm TeV})$ lower bounds for soft SUSY breaking parameters by PTA GW data, it is interesting to survey numerically if the parameter regions allowed by the muon $g-2$ explanation in $Z_3$-NMSSM can be consistent with the DW explanation of such stochastic GW background signals.

\subsection{Numerical results}

 We calculated numerically the tension of DWs, generated from the spontaneously discrete symmetry breaking in the $Z_3$-invariant NMSSM, within the parameters regions allowed by low-energy experimental data, including the following constraints other than those already encoded in the NMSSMTools\_5.6.2 package
 \bit
 \item[(i)] The CP-even component $S_2$ dominated combination of $H_u$ and $H_d$ doublets corresponds to the SM Higgs, which should lie in the combined mass range for the Higgs boson, $122 {\rm GeV}<M_h <128 {\rm GeV}$~\cite{ATLAS:higgs,CMS:higgs}. We adopt the uncertainty  $\pm 3$ {\rm GeV} instead of default 2 GeV because large $\lambda$ may induce additional ${\cal O}(1)$ GeV correction to $M_h$ at the  two-loop level~\cite{NMSSM:higgs2loop}.

  \item[(ii)] Bounds for low mass and high mass resonances at LEP, Tevatron, and LHC are taken into account by using the package HiggsBounds-5.5.0~\cite{higgsbounds511}and HiggsSignal-2.3.0~\cite{Bechtle:2013xfa}.

  \item[(iii)] Updated LHC exclusion bounds~\cite{ATLAS:2017kyf,CMS:2017arv,ATLAS:2017vjw}
  for sparticle searches and the lower mass bounds of charginos and sleptons from the LEP~\cite{LEPmass}. All relevant EW SUSY searches are taken into account, via CheckMATE2~\cite{Drees:2013wra,Kim:2015wza,Dercks:2016npn}.
 We discard the parameter points whose $R$ values obtained from the CheckMATE2.0 are larger than 1, i.e., excluded at $95\%$ CL.
  \item[(iv)] Constraints from B physics, such as $B \to X_s \gamma$, $B_s \to \mu^+ \mu^-$and $B^+ \to \tau^+ \nu_\tau$, etc~\cite{BaBar:2012fqh,BaBar:2012obs,LHCb-BsMuMu,Btaunu}.
  \item[(v)]  Vacuum stability bounds on the soft SUSY breaking parameters, including the semianalytic bounds for nonexistence of a deeper charge/color breaking minimum~\cite{Kitahara:2013lfa} and/or a metastable EW vacuum with a tunneling lifetime longer than the age of the Universe~\cite{vs2}.
  \item [(vi)] The relic abundance of the dark matter~(DM) should be below the upper bounds by the Planck result $\Omega_{DM} h^2= 0.1199\pm 0.0027$ \cite{Planck} in combination with the WMAP data~\cite{WMAP} (with a $10\%$ theoretical uncertainty). Besides, we require that the Spin-Independent~(SI) and Spin-Dependent~(SD) DM direct detection constraints, for~example, the~LUX~\cite{LUX:2016ggv}, XENON1T~\cite{XENON:2018voc,XENON:2019rxp}, and PandaX-4T~\cite{PandaX-4T:2021bab,PandaX:2022xas}--should be satisfied. We should note that such DM constraints can in fact be relaxed in gauge mediation scenarios, as the gravitino mass can be as light as ${\cal O}(1) {\rm eV}$, which can account for the present DM relic abundance for $m_{3/2}\sim {\rm keV}$. In our conservative survey, such DM constraints are included in our numerical results.


\item [(v)] The muon $g-2$ anomaly needs to be explained within the $2\sigma$ range for the recent updated combined data. We require the new physics contributions to $\Delta a_\mu$ to lie within
    \beqa
    \Delta a_\mu^{NP}\in [15.3, 34.5]\tm 10^{-10}~.
    \eeqa
\eit

We use NMSSMTools\_5.6.2~\cite{NMSSMTOOLS,Allanach:2008qq} to numerically scan the parameter spaces for muon the $g-2$ explanation with the following low-energy NMSSM inputs
\beqa
&& 0<\lambda<0.2,~~-0.2<\kappa<0.2,~~20<\tan\beta<60,~~-1000<\mu_{eff}<1000\mathrm{GeV}\nn \\
&&  -1000\mathrm{GeV}<A_{\lambda}<1000\mathrm{GeV},~~-500\mathrm{GeV}<A_{\kappa}<500\mathrm{GeV},\nn\\
&&|M_{1}|< 1000\mathrm{GeV},~~70\mathrm{GeV}<M_{2}<2000\mathrm{GeV},~~M_{3}=10\mathrm{TeV},\nn\\
&& 1000\mathrm{GeV}<M_{L_{3}}<3000\mathrm{GeV},~~1400\mathrm{GeV}<M_{E_{3}}<3000\mathrm{GeV},~~|A_{E_{3}}|<2000\mathrm{GeV},\nn\\
&& M_{Q_{3}}=10\mathrm{TeV},~M_{U_{3}}=10\mathrm{TeV},~~M_{D_{3}}=10\mathrm{TeV},~~ A_{U_{3}}=10\mathrm{TeV},~~A_{D_{3}}=10\mathrm{TeV},\nn\\
&& 80\mathrm{GeV}<M_{L_{1;2}}<500\mathrm{GeV},~~ 80\mathrm{GeV}<M_{E_{1;2}}<400\mathrm{GeV},~~|A_{E_{1;2}}|<500\mathrm{GeV},\nn\\
&& M_{Q_{1;2}}=10\mathrm{TeV},~~M_{U_{1;2}}=10\mathrm{TeV},~~M_{D_{1;2}}=10\mathrm{TeV}~.
\eeqa
Instead of using the soft SUSY breaking parameters $m^2_{H_u}$, $m^2_{H_d}$ and $m^2_{S}$, one usually trades them for $m_Z$,~$\tan\beta$ and $\mu_{eff}=\lambda s$ by implementing the scalar potential minimization conditions. The squarks and gluino are chosen to be heavy to evade the stringent LHC constraints on the colored sparticles. The bias terms that trigger the collapse of DWs are related to the explicit $Z_3$ breaking $\chi m_{3/2}$ parameter. We check that the presence of small explicitly $Z_3$ violation bias terms with $\chi m_{3/2}\sim 10^{-12} {\rm eV}$ will not alter the $Z_3$ invariant NMSSM spectrum.

The present-day relic abundance of stochastic GW backgrounds can be calculated numerically with the tensions of the DWs and the related bias terms for each parameter point that satisfy the constraints from (i) to (v). The BBN and $t_{dom}$ constraints in~(\ref{constraints:DWt}) are also taken into account.

We have the following discussions for our numerical results

\bit
 \item The $\chi S^2$ case:

From our estimation, we adopt the following range of $\chi m_{3/2}$
\beqa
10^{-16}{\rm eV}<\chi m_{3/2}<10^{-12} {\rm eV},
\eeqa
in our numerical scan.

\begin{figure}[hbt]
\begin{center}
\includegraphics[width=2.7in]{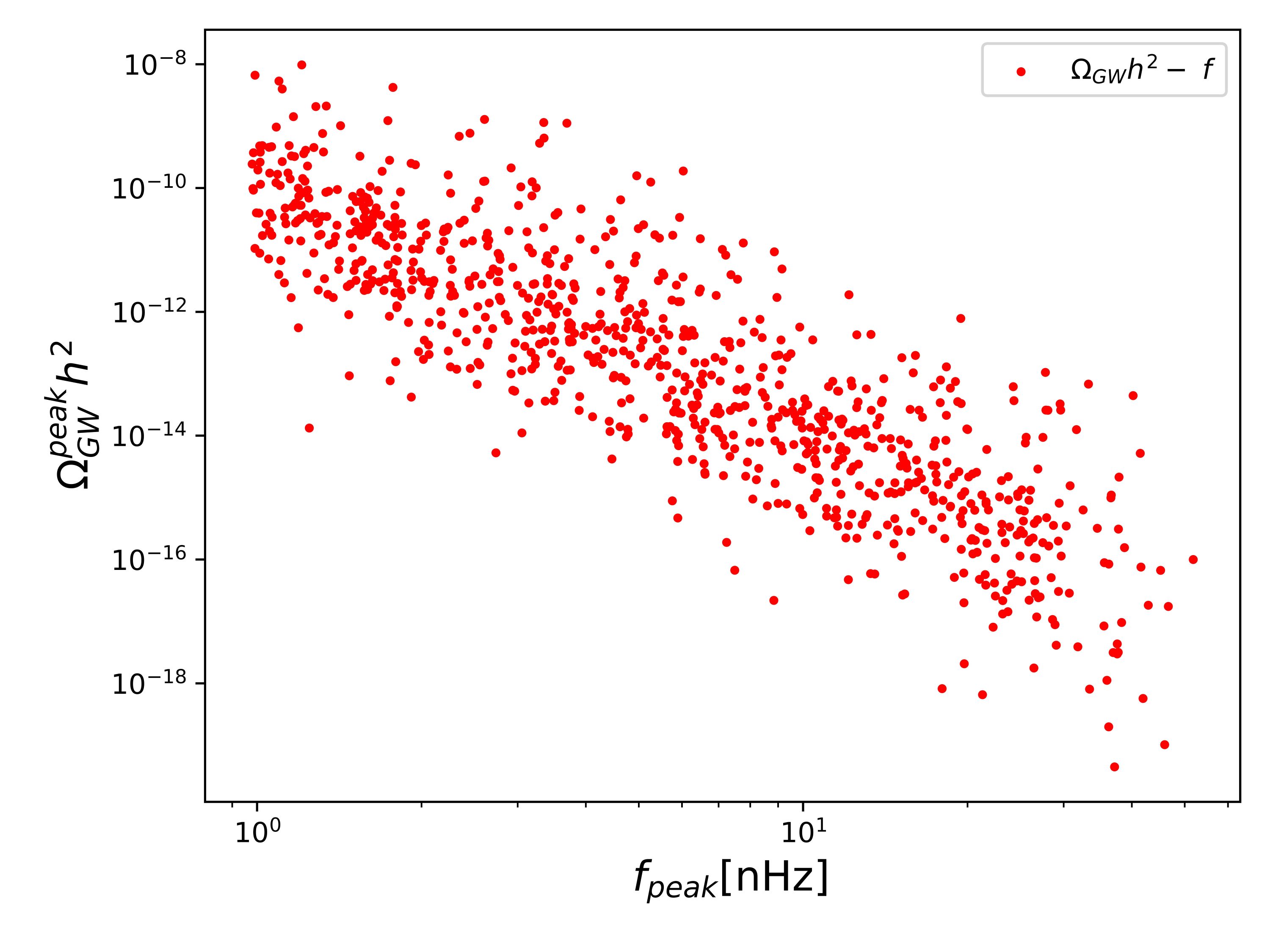}
\includegraphics[width=2.7in]{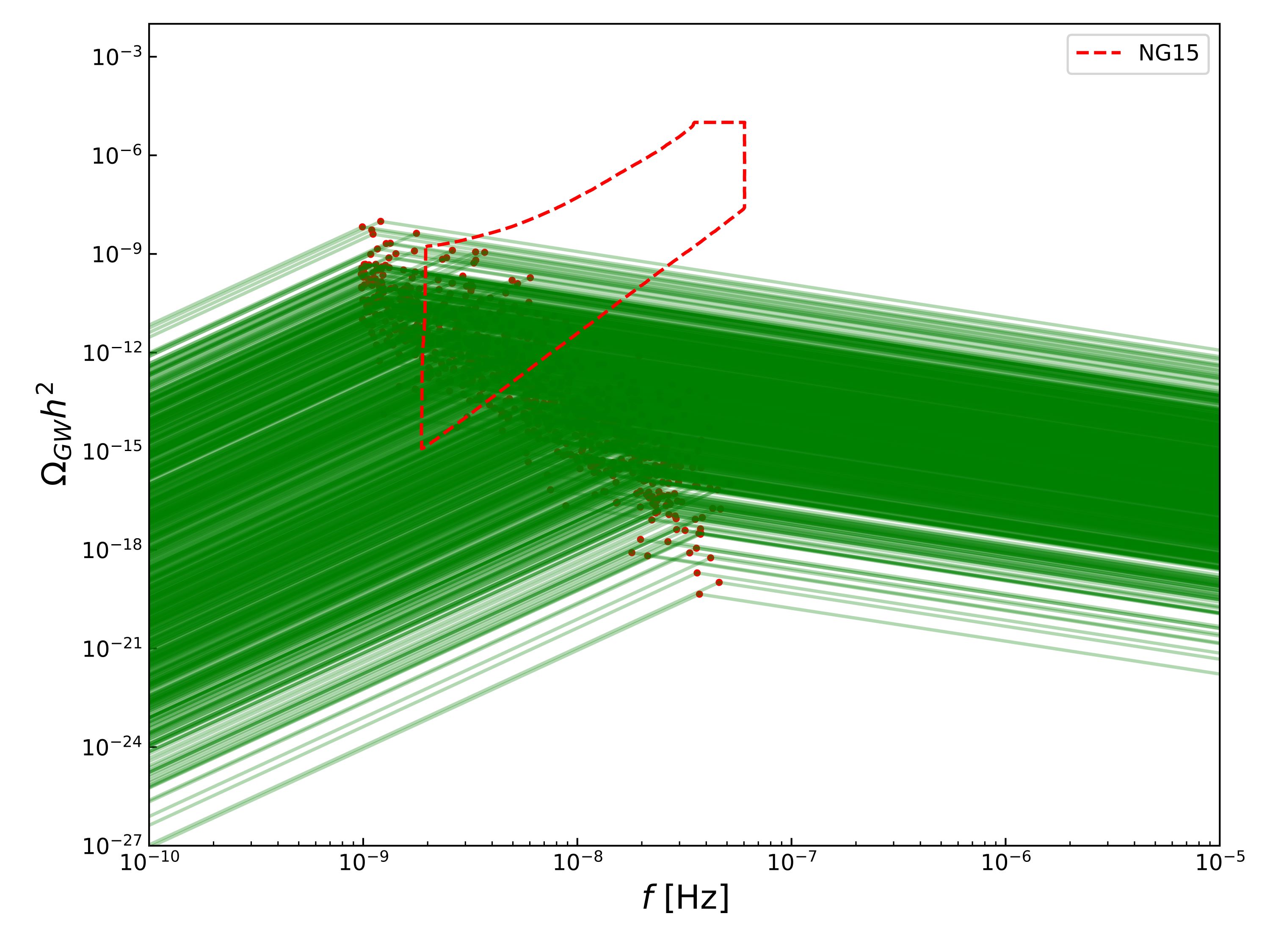}\\
\vspace{-.5cm}\end{center}
\caption{The left panel shows the present-day peak amplitude of stochastic GW background $\Omega_{gw}h^2$ versus the peak frequency $f_{peak}$ for the parameter points of $\chi S^2$ case. The corresponding GW spectra (with the NANOGrav observation data denoted by the red contour) are shown in the right panel. Each parameter point satisfies the experimental constraints (i)-(v). }
\label{fig1}
\end{figure}

\begin{figure}[hbt]
\begin{center}
\includegraphics[width=2.7in]{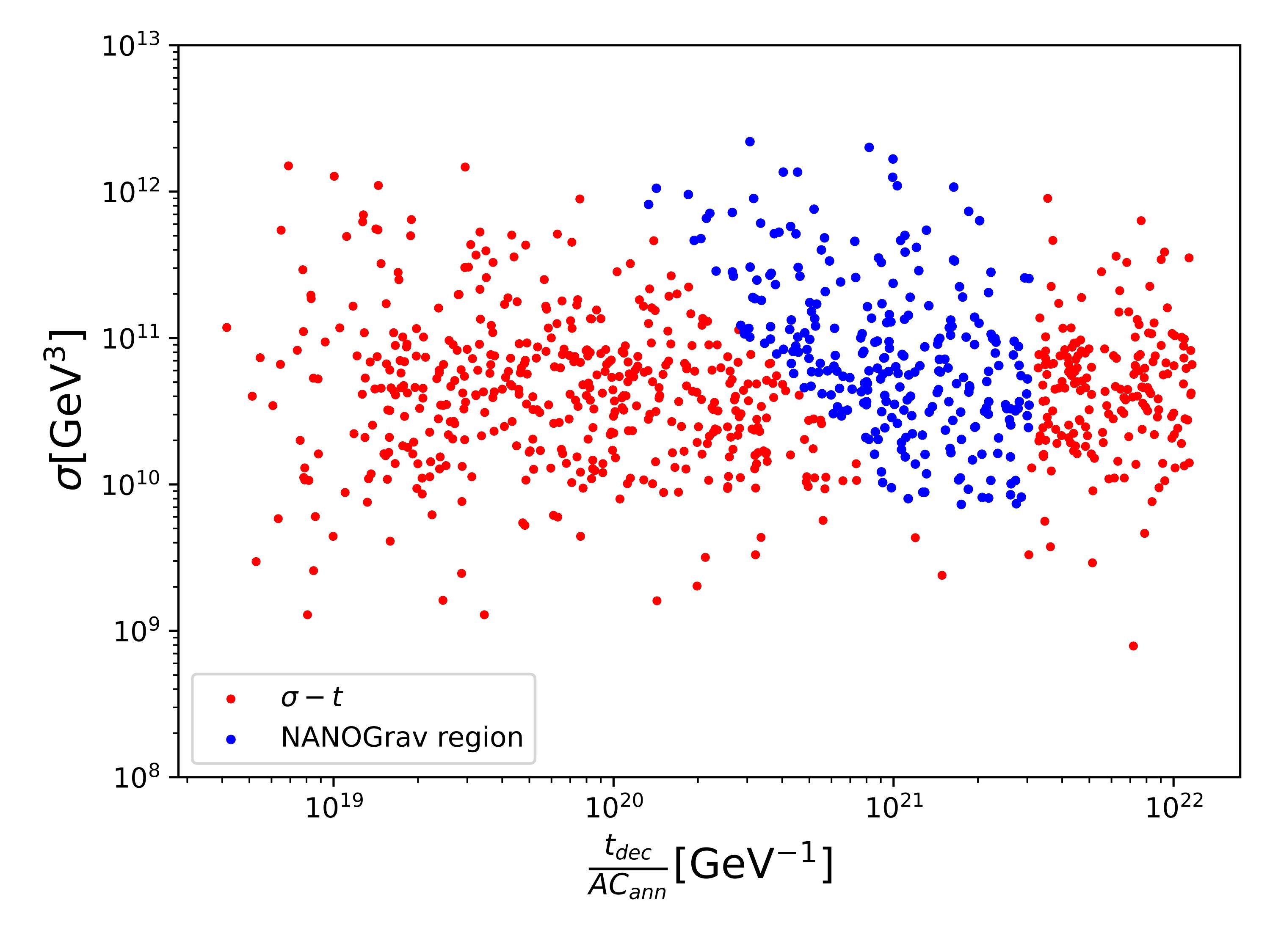}
\includegraphics[width=2.7in]{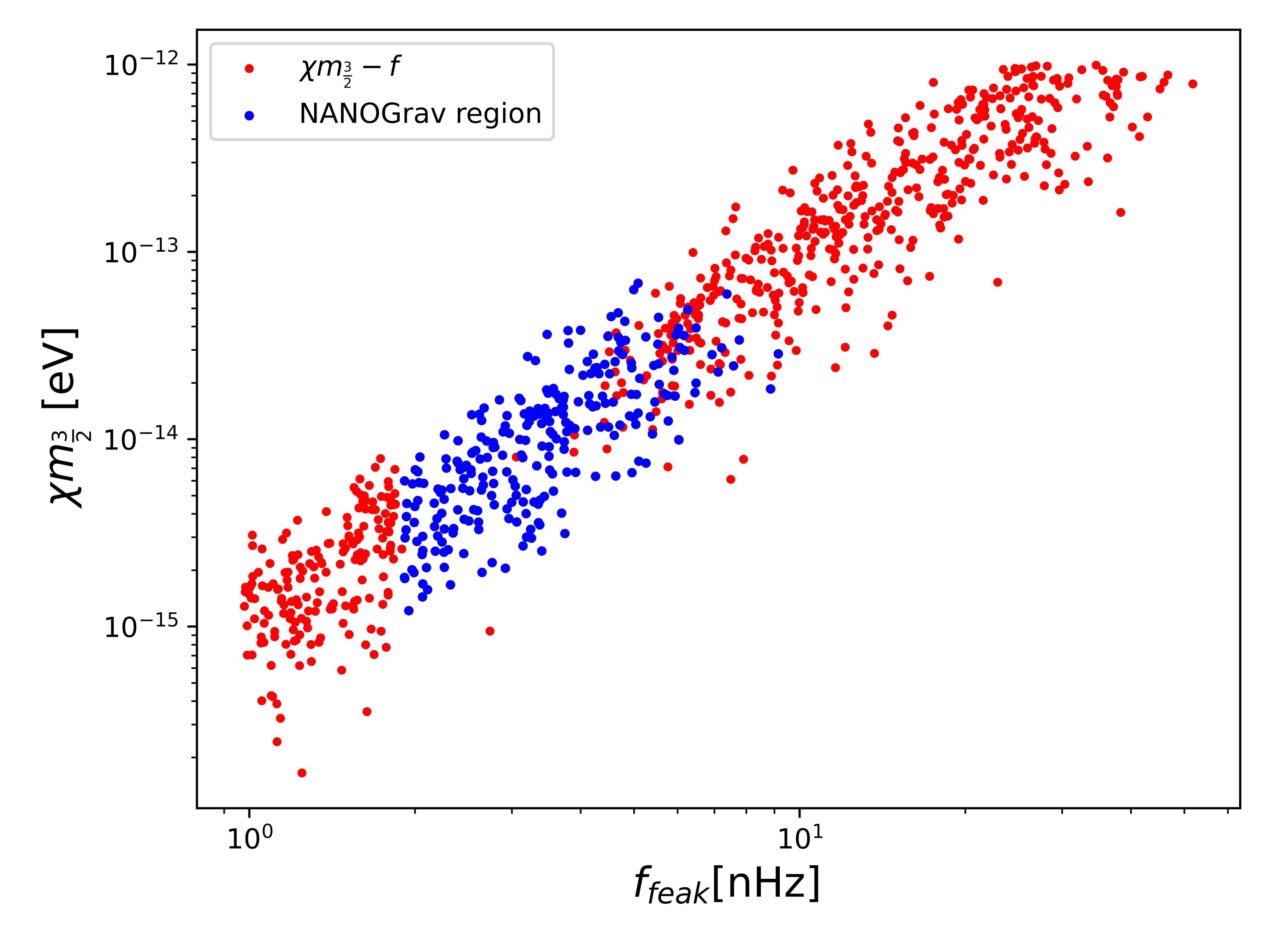}\\
\vspace{-.5cm}\end{center}
\caption{The tensions of the DWs versus their collapse time for the parameter points of $\chi S^2$ case are shown in the left panel. The parameter points that can account for the NANOGrav GW background observations are marked with blue color. Each parameter point satisfies the experimental constraints (i)-(v). The values of $\chi m_{3/2}$ versus the peak frequencies are shown in the right panel. }
\label{fig2}
\end{figure}
   The peak amplitudes of the GW backgrounds at the present time versus the corresponding peak frequencies are shown in the left panel of~Fig.\ref{fig1}. The corresponding GW spectra (with the NANOGrav observation data denoted by the red contour) are shown in the right panel of~Fig.\ref{fig1}. We can see that the GW spectrum for many of the survived parameter points can have overlap with the reconstructed posterior distributions for the NANOGrav observations of $nHz$ GWs so as that the DW explanation from $Z_3$-NMSSM is viable.

  We show in the left panel of Fig.\ref{fig2} our numerical results on the tensions of the DWs versus their collapse time. The parameter points, whose GWs spectra can overlap with the NANOGrav signal region, are marked with the color blue. From the distributions of the blue points, we can see that the PTA data can impose stringent constraints for the muon $g-2$ favored parameter space. As discussed in the previous section, the lower bounds from $\sigma$ (hence giving lower bounds for $m_{soft}$) can exclude some regions that can explain the muon $g-2$ with smaller $m_{soft}$. From the right panel of Fig.\ref{fig2}, we can see that the range of $\chi m_{3/2}$ that can account for the NANOGrav $nHZ$ GW background data should lie within $10^{-15}{\rm eV}\sim 10^{-13} {\rm eV}$. As anticipated, there is approximate linear dependence between $\chi m_{3/2}$ and $f_{peak}$.

\begin{figure}[hbt]
\begin{center}
\includegraphics[width=2.7in]{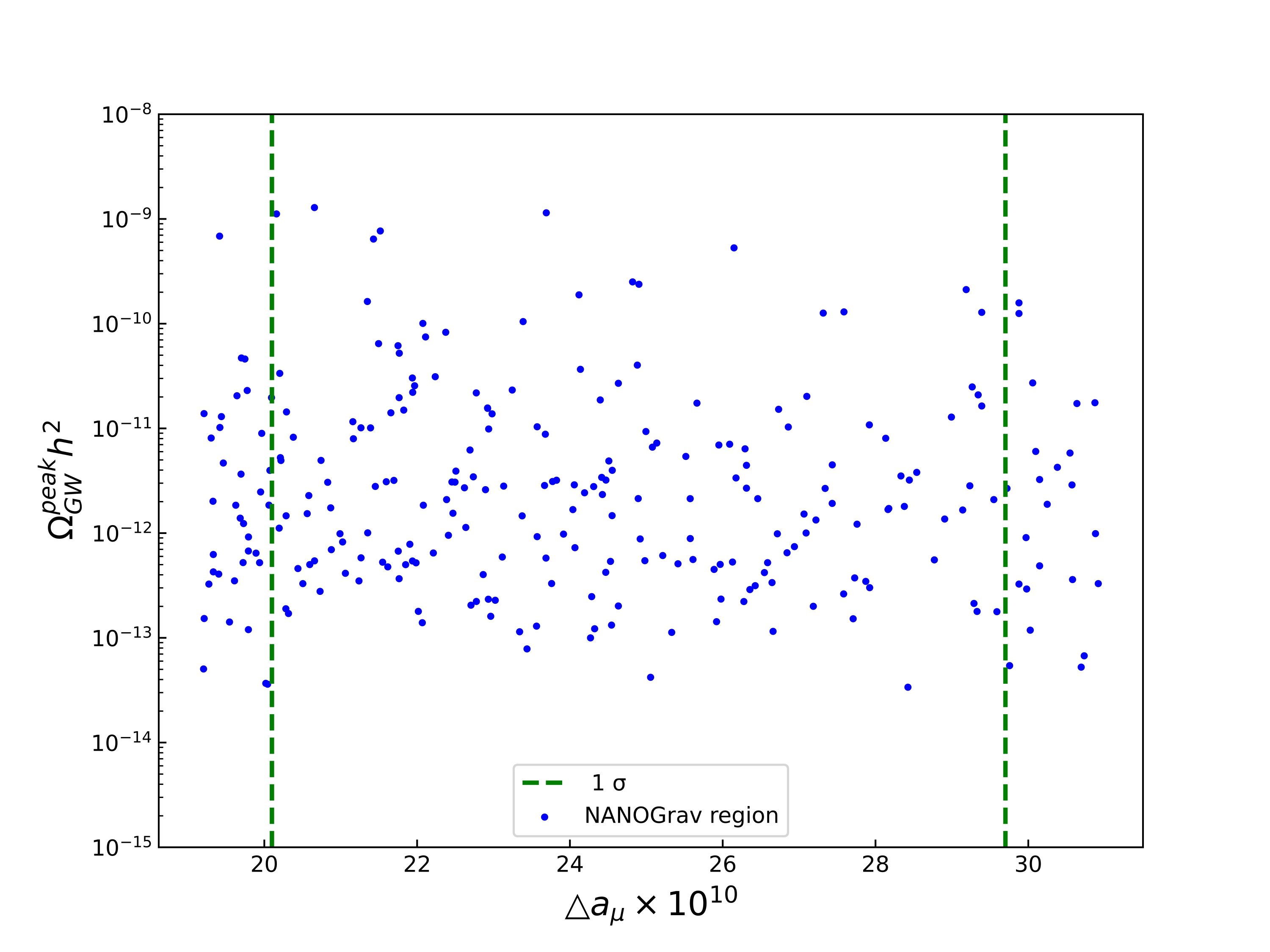}
\includegraphics[width=2.7in]{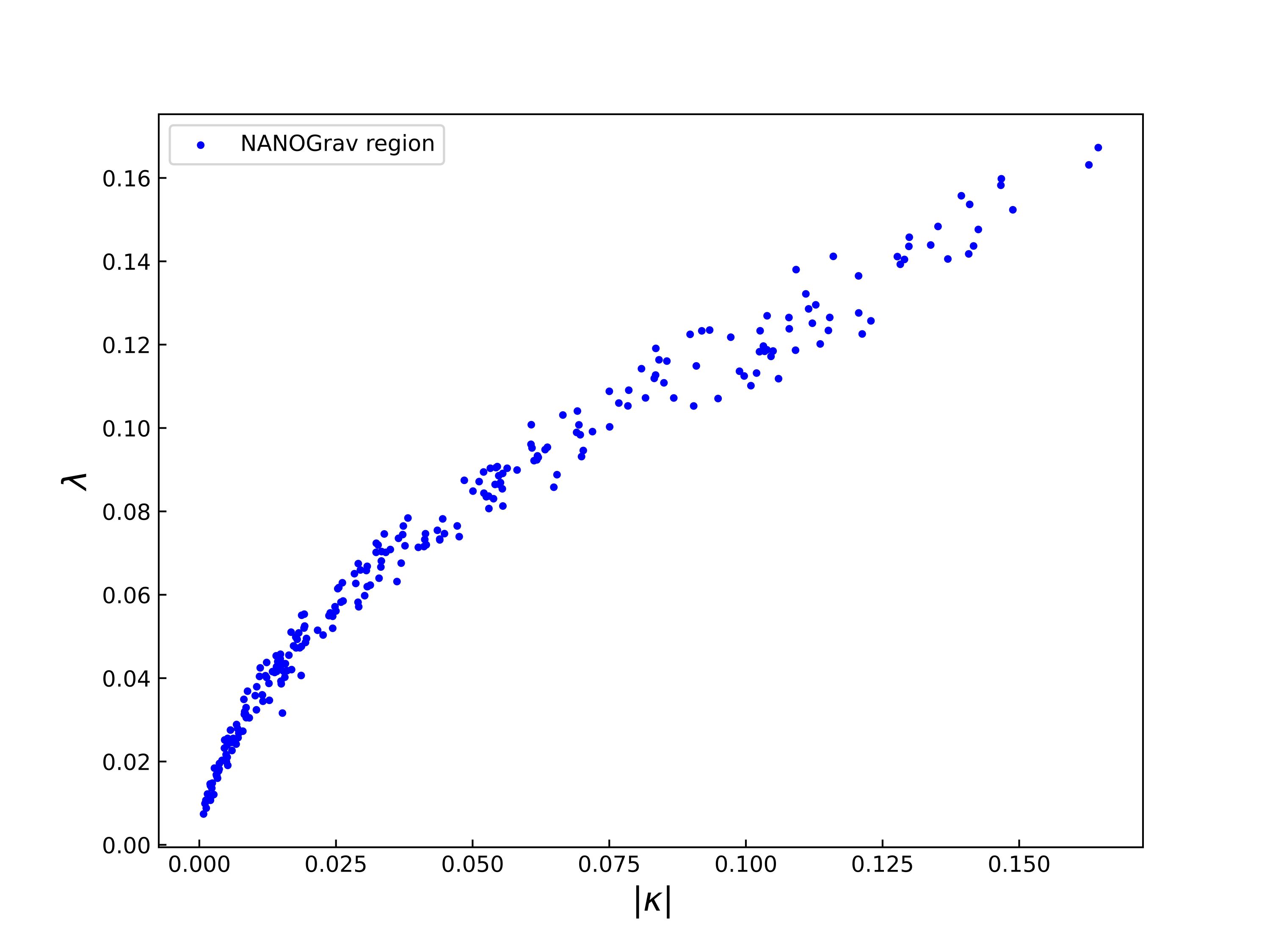}\\
\vspace{-.5cm}\end{center}
\caption{The tensions of the DWs versus their collapse time for the parameter points of $\chi S^2$ case are shown in the left panel. The parameter points that can account for the NANOGrav GW background observations are marked with the color blue. Each parameter point satisfies the experimental constraints (i)-(v). The values of $\chi m_{3/2}$ versus the peak frequencies are shown in the right panel. }
\label{fig3}
\end{figure}

  We show in the left panel of Fig.\ref{fig3} the SUSY contributions to muon $g-2$, the values $\Delta a_\mu$, versus the present-day peak amplitudes of stochastic GW background $\Omega_{gw}h^2$ for those survived points whose GWs spectra can overlap with the NANOGrav signal region. We can see that there are still large parameter regions that can explain the muon $g-2$ anomaly within $1\sigma$ range.
  Large values of $\Omega_{gw}h^2$ always favor large $\sigma$, hence large $m_{soft}$. As the SUSY explanation of $\Delta a_\mu$ in general favor small $m_{soft}$ of order ${\cal O}(10^2)${\rm GeV},  the preferred ranges of $\Delta a_\mu$ that can account for the muon $g-2$ anomaly favor low $\Omega_{gw}h^2$. The ranges of the $\lambda$ and $\kappa$ parameters of NMSSM that can account for the NANOGrav GW background observations are shown in the right panel of Fig.\ref{fig3}. Both of them are upper bounded to be be less than $0.18$.

\item The $\chi H_u H_d$ case:
\begin{figure}[hbt]
\begin{center}
\includegraphics[width=2.7in]{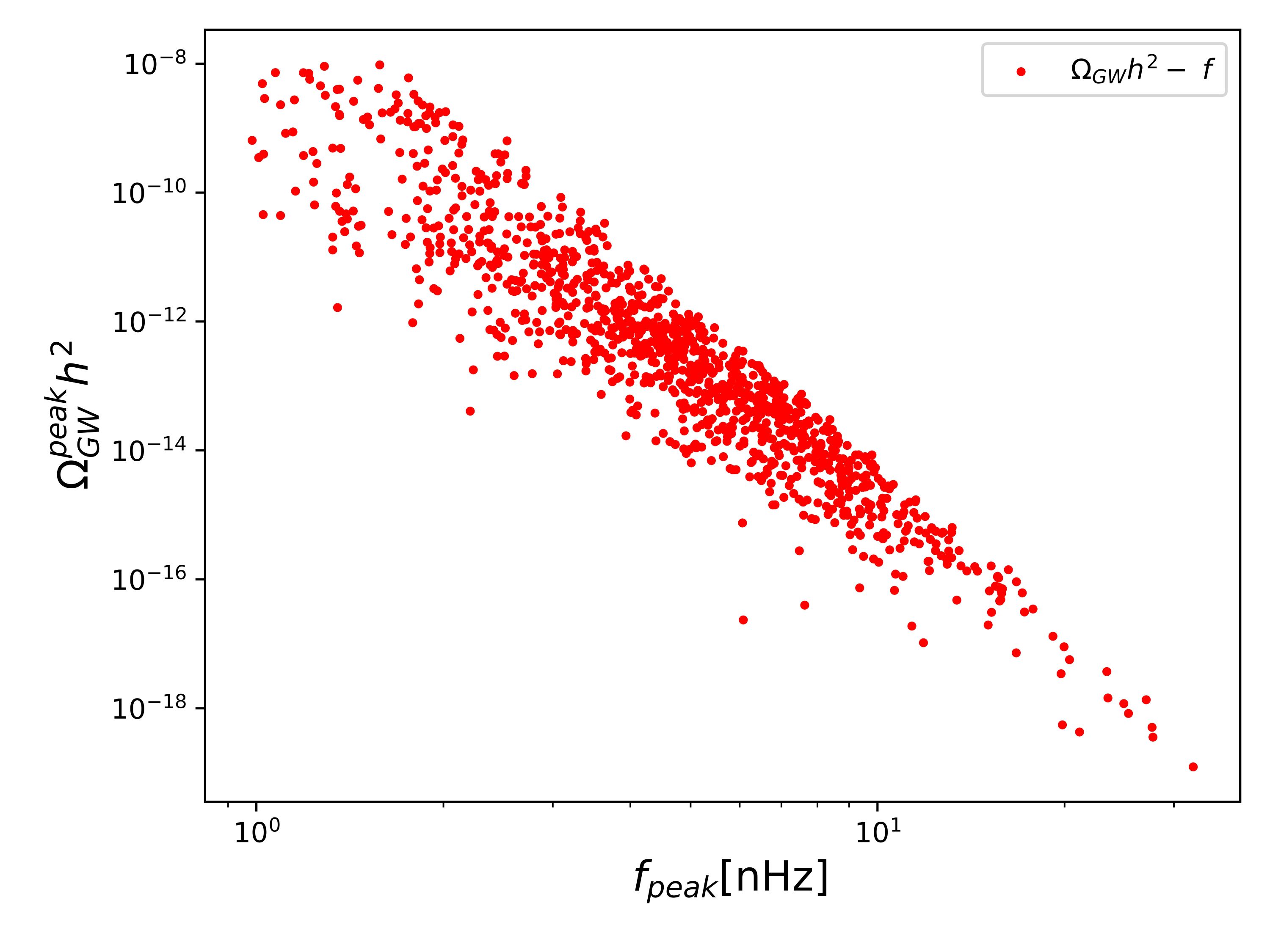}
\includegraphics[width=2.7in]{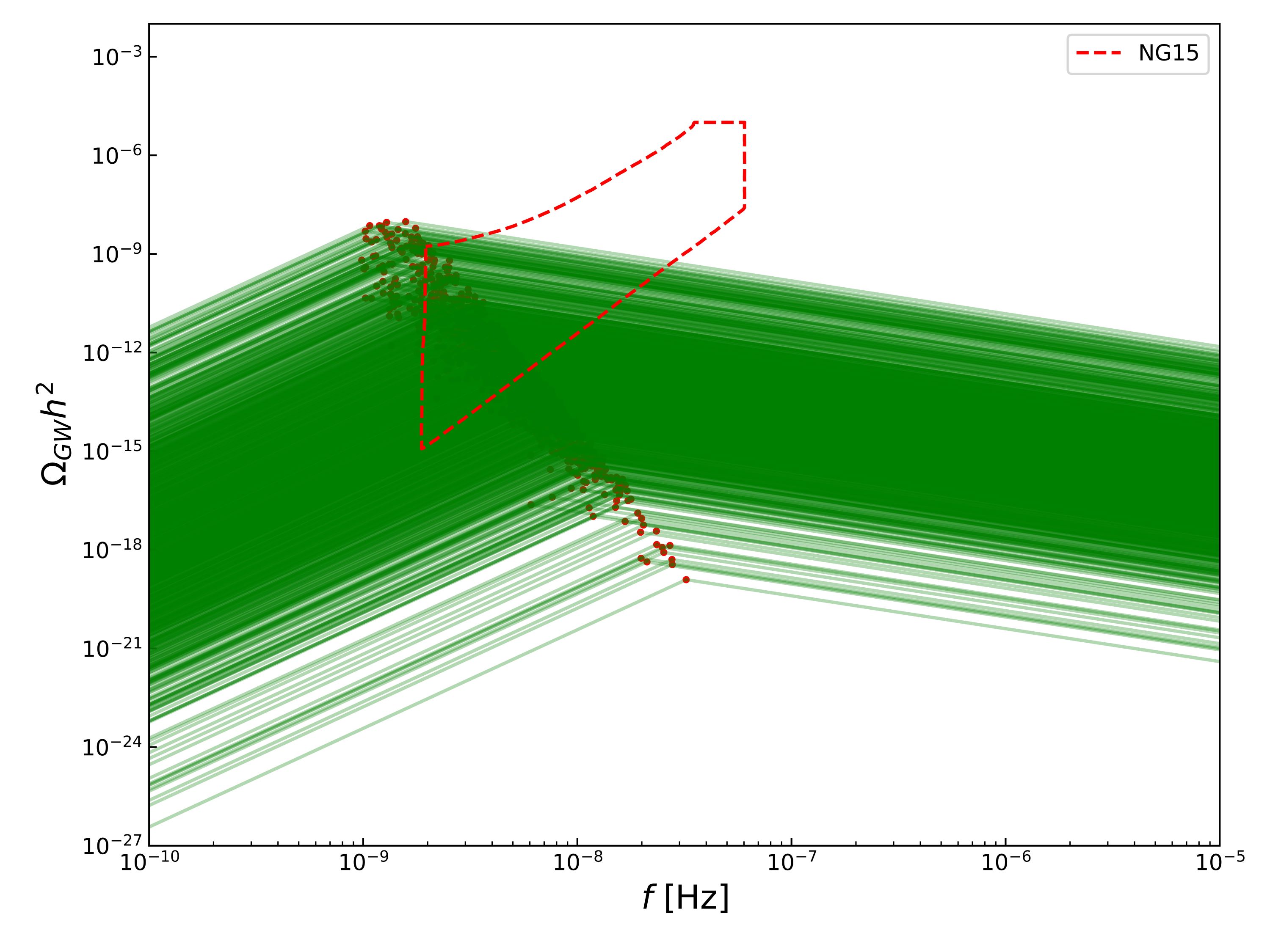}\\
\vspace{-.5cm}\end{center}
\caption{The left panel shows the present-day peak amplitude of stochastic GW background $\Omega_{gw}h^2$ versus the peak frequency $f_{peak}$ for the parameter points of $\chi H_uH_d$ case. The corresponding GW spectra (with the NANOGrav observation data denoted by the red contour) are shown in the right panel. Each parameter point satisfies the experimental constraints (i)-(v). }
\label{fig4}
\end{figure}

From our estimation, we adopt the following range of $\chi m_{3/2}$
\beqa
10^{-13}{\rm eV}<\chi m_{3/2}<10^{-9} {\rm eV},
\eeqa
in our numerical scan.

We can have similar discussions for this case. The peak amplitudes of the GW backgrounds at the present time versus the corresponding peak frequencies is shown in the left panel of~Fig.\ref{fig4}. The corresponding GW spectra are shown in the right panel of~Fig.\ref{fig4}. Again, the GW spectrum for many the surviving parameter points can have overlap with the reconstructed posterior distributions for the NANOGrav observations of $nHz$ GWs. So, it is obvious that the collapse of DWs from approximate $Z_3$-invariant NMSSM with the small explicitly $Z_3$ breaking $\chi H_uH_d$ term (and its soft SUSY breaking B-terms) can explain the PTA observations.

\begin{figure}[hbt]
\begin{center}
\includegraphics[width=2.7in]{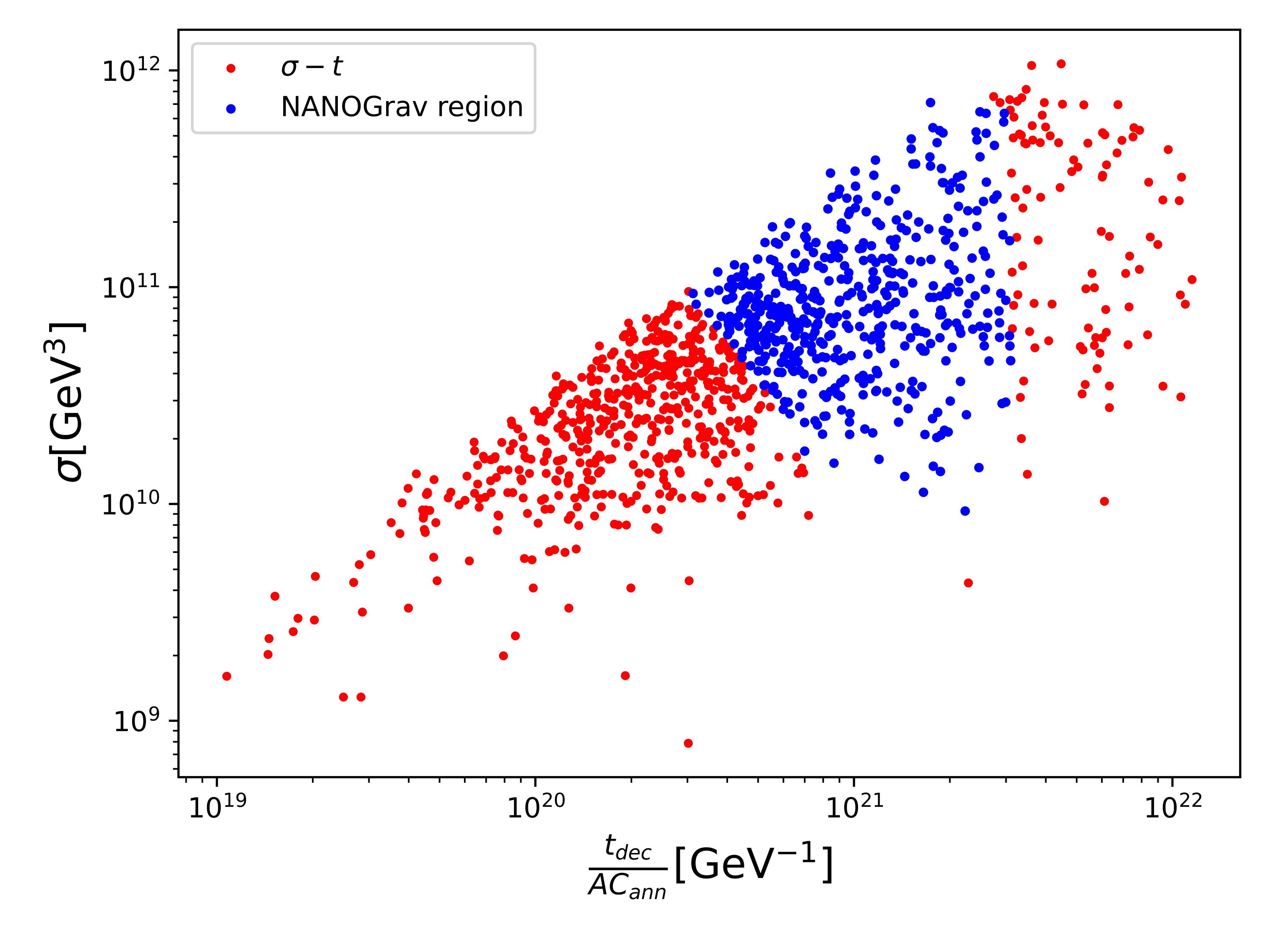}
\includegraphics[width=2.7in]{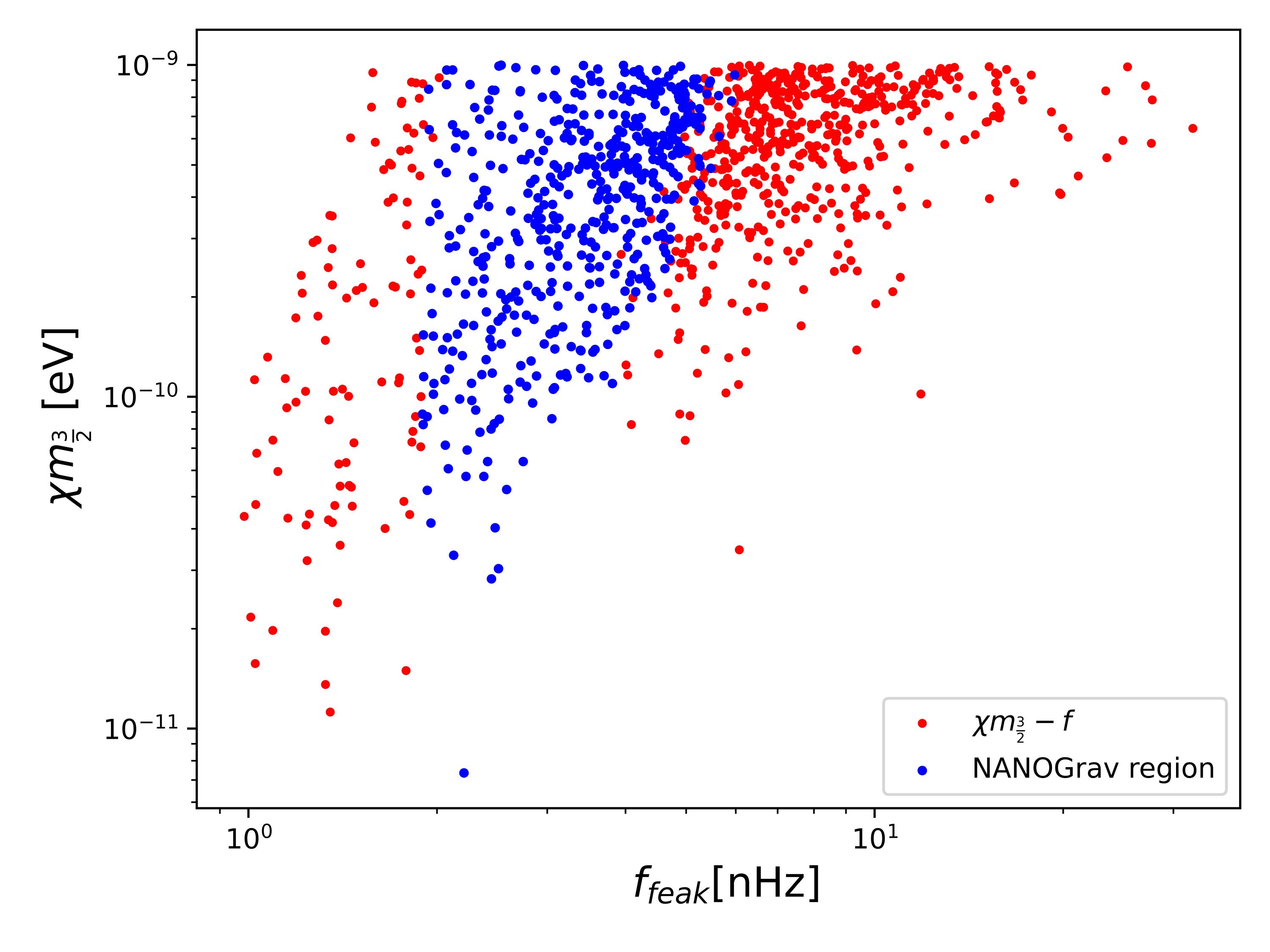}\\
\vspace{-.5cm}\end{center}
\caption{The tensions of the DWs versus their collapse time for the parameter points of $\chi H_u H_d$ case are shown in the left panel. The parameter points that can account for the NANOGrav GW background observations are marked with the color blue. Each parameter point satisfies the experimental constraints (i)-(v). The values of $\chi m_{3/2}$ versus the peak frequencies are shown in the right panel. }
\label{fig5}
\end{figure}

  In the left panel of Fig.\ref{fig5}, we show our numerical results on the tensions of the DWs versus their collapse time. From the distributions of the blue points, which denote those parameter points whose GWs spectrum can be consistent with the NANOGrav signal region, we can see that the PTA data can again impose stringent constraints for the muon $g-2$ favored parameter space. Similarly, the lower bounds from $\sigma$ that lead to lower bounds for $m_{soft}$ can exclude some regions that can explain the muon $g-2$ with light sparticle masses. From the right panel of Fig.\ref{fig5}, we can see that the range of $\chi m_{3/2}$ that can account for the NANOGrav nHZ GW backgroun5d data should lie within $10^{-11}{\rm eV}\sim 10^{-9} {\rm eV}$. The correlations between $\chi m_{3/2}$ and $f_{\rm peak}$ are not obviously linear, as the bias terms in this case can differ in a relatively wide range, which involves $v^2$,$\tan\beta$ and $m_{soft}$.

\begin{figure}[hbt]
\begin{center}
\includegraphics[width=2.7in]{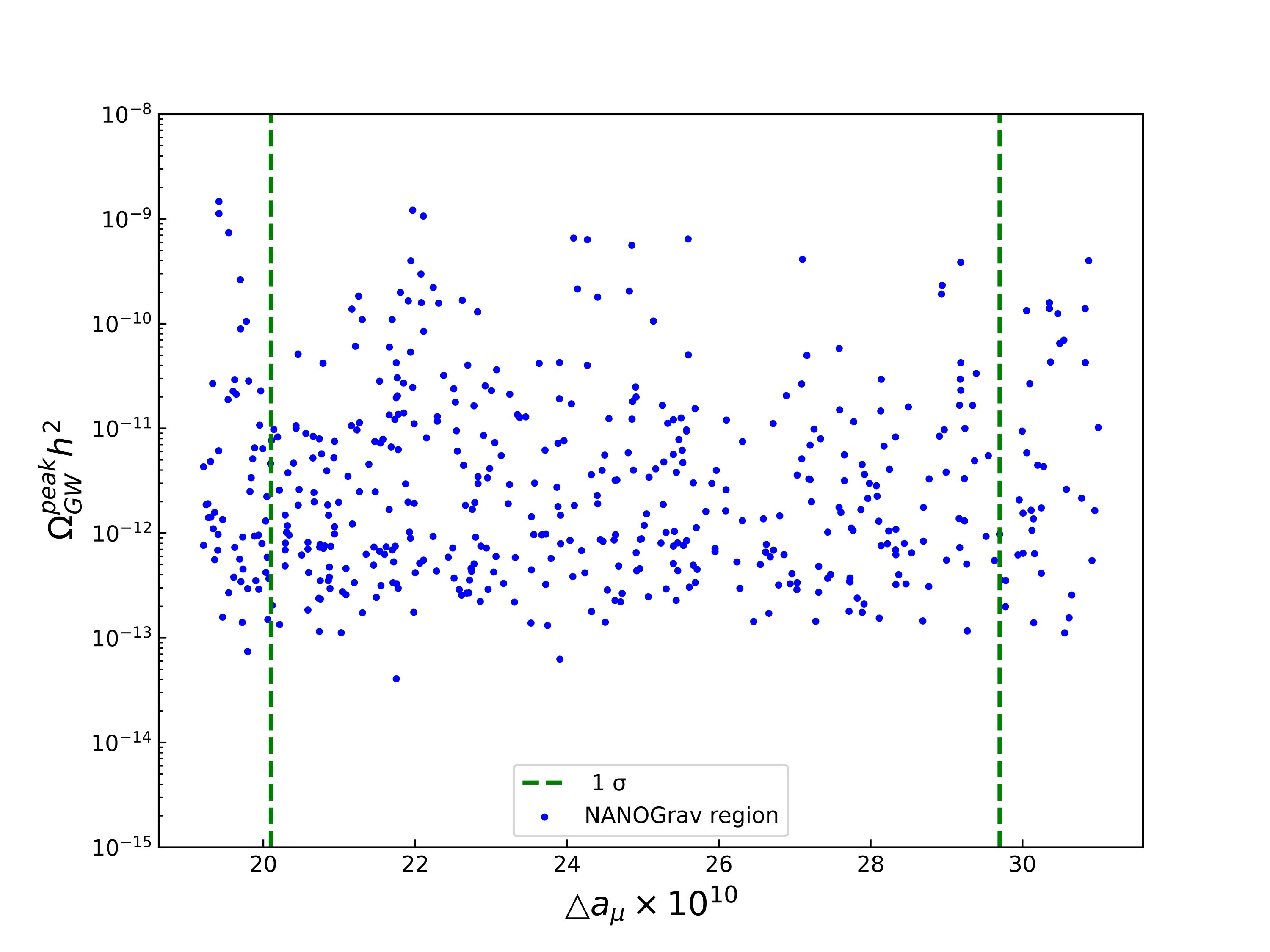}
\includegraphics[width=2.7in]{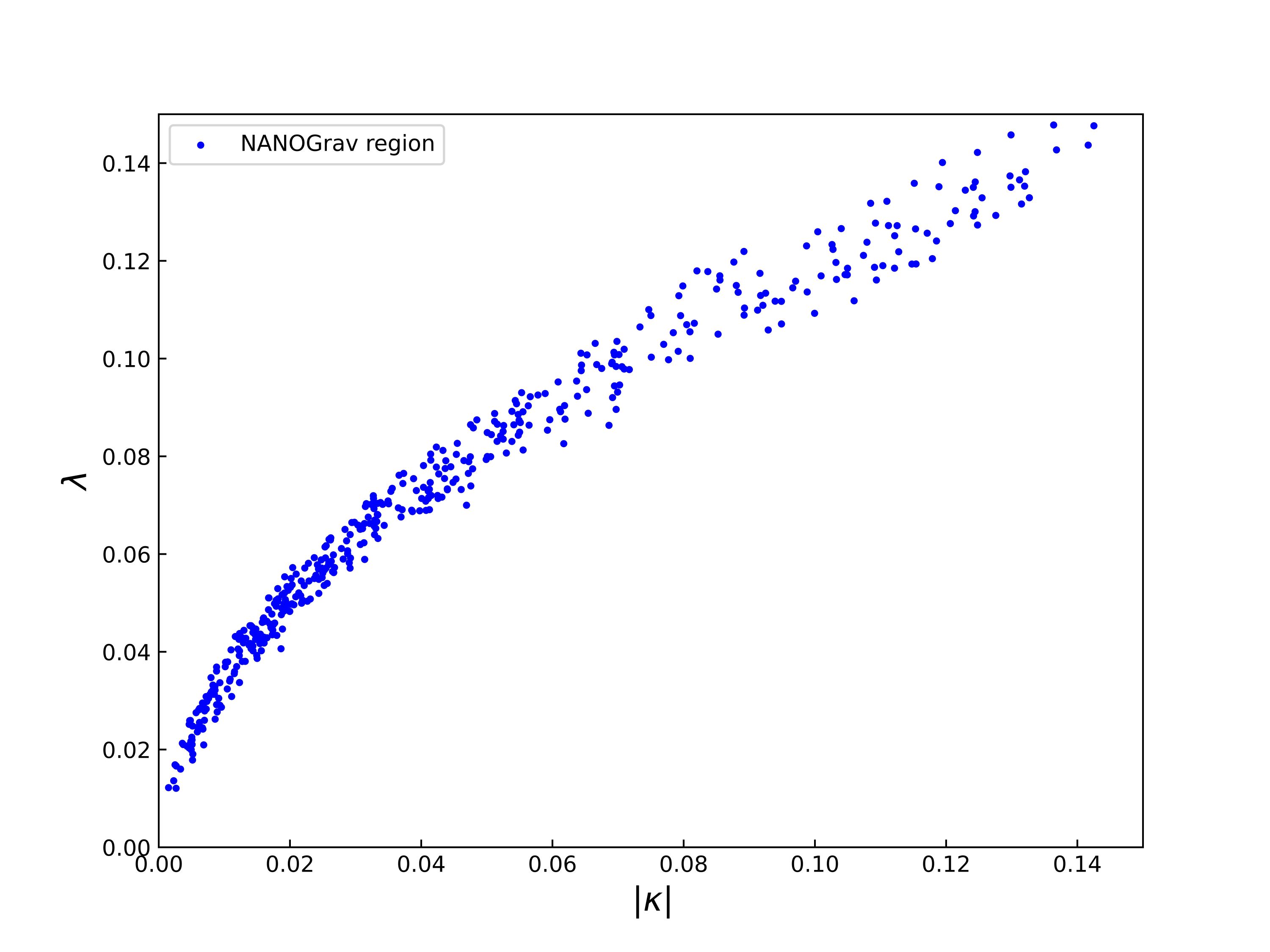}\\
\vspace{-.5cm}\end{center}
\caption{The tensions of the DWs versus their collapse time for the parameter points of $\chi H_uH_d$ case are shown in the left panel. The parameter points that can account for the NANOGrav GW background observations are marked with the color blue. Each parameter point satisfies the experimental constraints (i)-(v). The values of $\chi m_{3/2}$ versus the peak frequencies are shown in the right panel. }
\label{fig6}
\end{figure}

  We show in the left panel of Fig.\ref{fig6} the total SUSY contributions to muon $g-2$, the value $\Delta a_\mu$, versus the present-day peak amplitudes of stochastic GW background $\Omega_{gw}h^2$ for those survived points whose GWs spectra can overlap with the NANOGrav observations. Again, there are still large parameter regions that can explain the muon $g-2$ anomaly within $1\sigma$ range. The NANOGrav GW background data can set upper bounds for $\lambda$ and $\kappa$ parameters of NMSSM, which should be less than $0.14$ (see the right panel of Fig.\ref{fig6}).

 \eit
\section{\label{sec:conclusions} Conclusions}
The spontaneous breaking of the discrete $Z_3$ symmetry in approximate $Z_3$-invariant Next-to Minimal Supersymmetric Standard Model by the VEVs of Higgs fields can lead to the formation of DWs in the Universe. Such potentially problematic DWs can collapse and lead to the stochastic GW background signals observed by PTA collaborations due to some explicitly $Z_3$ breaking terms in the NMSSM effective superpotential and scalar potential. In the presence of a hidden sector, such terms may originate from the geometric superconformal breaking with holomorphic quadratic correction to frame function when the global scale-invariant superpotential is naturally embedded into the canonical superconformal supergravity models. The smallness of such $Z_3$ breaking mass parameters in the NMSSM may be traced back to the original superconformal invariance.

Naive estimations indicate that SUSY explanation to the muon $g-2$ anomaly can have tension with the
constraints on SUSY by PTA data, because large SUSY contributions to $\Delta a_\mu$ in general
need relatively light superpartners of order ${\cal O}(10^2) {\rm GeV}$ while the lower bounds for present $\Omega_{gw}^0$ can also set the lower bounds for the tension of DWs $\sigma$, leading approximately to ${\cal O}(1) {\rm TeV}$ lower bound for $m_{soft}$ with the BBN upper bounds for the decay time of DWs.  We calculate numerically the signatures of GWs produced from the collapse of DWs by the corresponding bias terms. We find that the observed nHZ stochastic GW background by NANOGrav, etc., can indeed be explained with proper tiny values of $\chi m_{3/2}\sim 10^{-14}{\rm  eV}$ for $\chi S^2$ case (and $\chi m_{3/2}\sim 10^{-10}{\rm  eV}$ for $\chi H_u H_d$ case), respectively. Such tiny values can be natural consequences of higher-dimensional operators in the Kahler potential with $U(1)_R$ symmetry, which do not have the notorious tadpole problem. Besides, there are still some parameter points--whose GWs spectra intersect with the NANOGrav signal region, can explain the muon $g-2$ anomaly to $1\sigma$ range.

\begin{acknowledgments}
We are very grateful to the referees for good suggestions.  This work was supported by the
Natural Science Foundation of China under Grant No. 12075213 and 12335005 and by the Key Research Project of Henan Education Department for colleges and universities under Grant No. 21A140025.
\end{acknowledgments}

\end{document}